\documentclass[proof,linenumber]{WileyASNA-v1}

\articletype{Article Type}%

\received{26 April 2024}
\revised{15 May 2024}
\accepted{26 May 2024}

\begin{document}

\title{Spectroscopic and Dynamic Orbital Analyses of Metal-Poor and High Proper Motion Stars: I. HD\,8724 and HD\,195633}

\author[1]{M. Marışmak}

\author[2]{T. Şahin*}

\author[1]{F. Güney}

\author[3]{O. Plevne}

\author[3]{S. Bilir}

\authormark{Marışmak \textsc{et al.}}

\address[1]{\orgdiv{Institute of Graduate Studies in Science}, \orgname{Akdeniz University}, \orgaddress{\state{Antalya}, \country{Türkiye}}}

\address[2]{\orgdiv{Faculty of Science, Department of Space Sciences and Technologies}, \orgname{Akdeniz University}, \orgaddress{\state{Antalya}, \country{Türkiye}}}

\address[3]{\orgdiv{Faculty of Science, Department of Astronomy and Space Sciences}, \orgname{Istanbul University}, \orgaddress{\state{Istanbul}, \country{Türkiye}}}

\corres{*Timur Şahin, Faculty of Science, Department of Space Sciences and Technologies, Antalya, Türkiye.\\ \email{timursahin@akdeniz.edu.tr}}

%\presentaddress{This is sample for present address text this is sample for present address text}

\abstract{In this study, spectral, age, kinematic, and orbital dynamical analyses were conducted on metal-poor and high proper-motion (HPM) stars, HD\,8724 and HD\,195633, selected from the Solar neighborhood. This analysis combines detailed abundance measurements, kinematics, and orbital dynamics to determine their origin. Standard 1D local thermodynamic equilibrium analysis provides a fresh determination of the atmospheric parameters: $T_{\rm eff}=$4700$\pm$115 K, $\log g=$ 1.65$\pm$0.32 cgs, [Fe/H]=-1.59$\pm$0.04 dex, and a microturbulent velocity $\xi=$ 1.58$\pm$0.50 km s$^{\rm -1}$ for HD\,8724 and  $T_{\rm eff}=$6100$\pm$205 K, $\log g=$3.95$\pm$0.35 cgs, [Fe/H]=-0.52$\pm$0.05 dex, and $\xi=$1.26$\pm$0.50 km s$^{\rm -1}$ for HD\,195633. The ages were estimated using a Bayesian approach (12.25 Gyr for HD\,8724 and 8.15 Gyr for HD\,195633). The escape scenarios of these stars from 170 candidate globular clusters (GCs) in the Galaxy were also investigated because of their chemical and physical differences (HPM and metal-poor nature). Accordingly, the calculated probability of encounter ($59\%$) for HD\,8724 at a distance of five tidal radius suggests that star HD\,8724 may have escaped from NGC\,5139 ($\omega$ Cen), supported by its highly flattened orbit and may belong to a sub-population of this GC. Conversely, HD\,195633's kinematics, age, and metal abundances point towards an escape from the bulge GC NGC\,6356.}

\keywords{(Galaxy:) solar neighborhood, Galaxy: halo, stars: abundances, stars: kinematics and dynamics}

%\jnlcitation{\cname{%
%\author{Williams K.}, 
%\author{B. Hoskins}, 
%\author{R. Lee}, 
%\author{G. Masato}, and 
%\author{T. Woollings}} (\cyear{2016}), 
%\ctitle{A regime analysis of Atlantic winter jet variability applied to evaluate HadGEM3-GC2}, \cjournal{Q.J.R. Meteorol. Soc.}, \cvol{2017;00:1--6}.}

\fundingInfo{Türkiye Bilimsel ve Teknolojik Araştırma Kurumu, Grant/Award Number:119F072}

\maketitle

%\footnotetext{\textbf{Abbreviations:} ANA, anti-nuclear antibodies; APC, antigen-presenting cells; IRF, interferon regulatory factor}

\section{Introduction}\label{sec1}

\label{introduction}
Metal-poor stars contain a fossil record of the chemical composition of the Milky Way, allowing the study of the formation of the Galaxy and the early stages of chemical evolution. By studying the chemical abundances in metal-poor stars, astronomers can infer the processes of nucleosynthesis that occurred shortly after the Big Bang and trace the chemical evolution of the Milky Way \citep[e.g.,][]{Mishenina2024, klessen2023,bensby2014}. Additionally, the distribution and frequency of metal-poor stars across the Galaxy offer insights into the formation and accretion history of the Galactic halo \citep{frebel2015}. 

Extensive studies, such as those by \citet{cayrel2004}, \citet{beers2005}, and \citet{norris2013a}, have investigated the chemical abundance of these stars, revealing significant insights into the processes of nucleosynthesis and the enrichment history of the Galaxy. Furthermore, metal-poor stars play an essential role in galactic archaeology, helping reconstruct the assembly history and chemical enrichment of their host galaxies \citep[e.g.,][]{bensby2003, bensby2005, bensby2014, norris2013b, karatas2005, karaali2019, nissen2011, plevne2020}.

Beyond their individual importance, metal-poor stars within globular clusters offer valuable constraints on the age and formation mechanisms of these ancient systems \citep{carretta2009a}. The advent of large-scale sky surveys have allowed for more detailed studies of these stars, categorizing them based on their metal abundances into various groups \citep{beers2005}. For instance, studies using data from ground-based large sky surveys such as the Apache Point Observatory Galactic Evolution Experiment (APOGEE; \citealp[] {majewski2017}; \citealp[]{prieto2008}), the Galactic ArchaeoLogy with HERMES project (GALAH; \citealp[]{desilva2015}), the Radial Velocity Experiment (RAVE; \citealp[]{steinmetz2006}), and Sloan Extension for Galactic Understanding and Exploration (SEGUE-1 and SEGUE-2, \citealp[]{yanny2009}; \citealp[]{eisenstein2011}) have significantly enhanced our understanding of different stellar populations within the Milky Way \citep{belokurov2007, kepley2007, sesar2007, klement2010, coskunoglu2012, bilir2012, helmi2017, li2019}.

The study of the spatial distributions of stars with different metal abundances contributes to the determination of the structural parameters of Galactic populations and helps to reveal the Milky Way  structure \citep{karaali2004, bilir2008}. Moreover, the comparison of the density profiles of stars observed by systematic sky surveys, such as the Sloan Digital Sky Survey \citep[SDSS, ][]{York2000}, with Galaxy models allows the determination of the percentage of metal-poor stars in the solar neighborhood \citep{ak2007a, ak2007b, bilir2006a, bilir2006b}. There are also metal-poor stars associated with the thick disc and halo in the Solar neighbourhood \citep{bensby2014, doner2023}. In fact, these thick disc stars in the Solar neighbourhood have been found to be richer in $\alpha$-elements than the thin disk \citep[e.g.,][]{fuhrmann1998, bensby2003, adibekyan2013, recioblanco2014}.

Notably, catalogs of high proper-motion stars have not only increased the number of known metal-poor stars \citep[see e.g.,][]{carney1996} but have also identified extremely metal-poor stars \citep[e.g.,][]{christlieb1998, matijevic2017}. \cite{sahin2020}, in addition to high-resolution spectroscopic analysis and afresh calculation of their Bayesian ages, utilized kinematic and dynamics orbital analysis of selected F-type stellar systems in the northern hemisphere of the Galaxy to provide an effective perspective that offered new insight into the physical nature of metal-poor systems, i.e., their population type and Galactic origin.  

%Figure 1
\begin{figure*}
    \centering
    \includegraphics[width=17.90cm,height=8.6cm,angle=0]{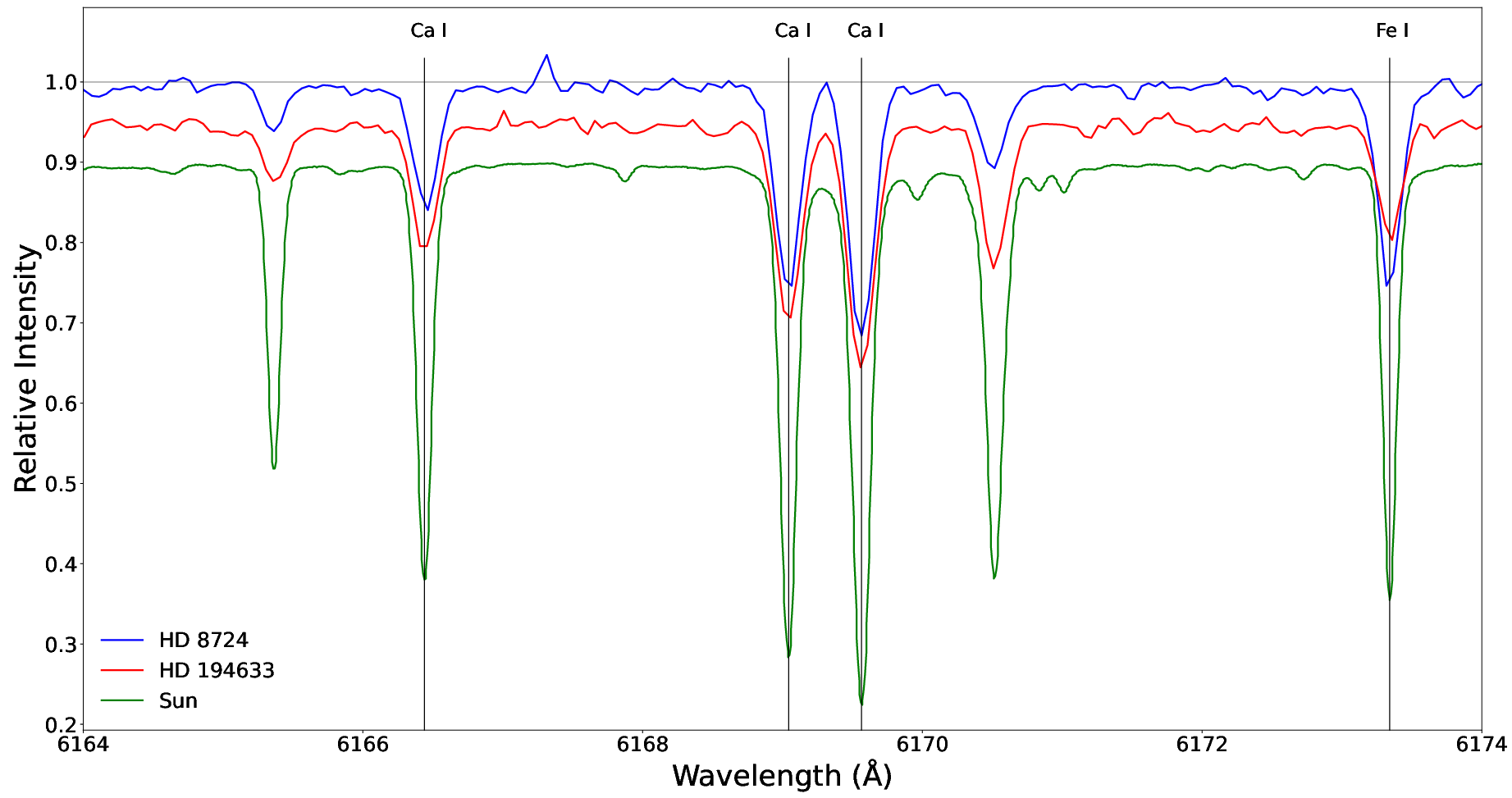}
    \caption{Small regions of the spectra for HD\,8724 (blue), HD\,195633 (red), and Sun (green). The spectra are vertically shifted for convenience. The neutral iron and calcium lines used in the analysis are indicated by a black vertical line, and the lines not used in the analysis are indicated by a red dashed line.}
    \label{fig:spec}
\end{figure*}

%Table 1
\begin{table*}
\setlength{\tabcolsep}{8.0pt}
\caption{Equatorial coordinates, spectral types, and {\sc ELODIE} spectral data were provided for HD\,8724 and HD\,195633, respectively. The corrected {\sc ELODIE} spectra accounted for the reported heliocentric radial velocities, which were cross-checked by the Solar spectrum.}
\label{table:1}
\centering
\begin{tabular}{lccccccc}
\hline
    Star       & $\alpha$ &$\delta$         &Spectral Type&Exposure     &$S/N^{*}$ & $V_{\rm Rad}$ & MJD       \\
                & (hh:mm:ss)  & (dd:mm:ss) &        &  (sec)        &    &  (km s$^{\rm -1}$)         & (2400000+)    \\
    \hline
    HD\,8724  & 01 26 17.60 & +17 07 35.12  & G5*  & 3600  & 102.2 & -113.52$\pm$1   &  50361.991141 \\
    HD\,195633  & 20 32 23.99 & +06 31 03.26  & G0V* & 3600  & 111.9 & ~~-45.94$\pm$1    & 50360.852275 \\
    
    \hline
\end{tabular}
\begin{minipage}{17cm}
*: $S/N$ values are reported in the {\sc ELODIE} spectra for ~5500 \AA ~wavelength. The spectral types of stars were obtained from the SIMBAD database \citep{Wenger20}.
\end{minipage}
\end{table*}

In this study, we present results from the spectral, kinematic, and dynamical orbital analyses of HD\,8724 and HD\,195633, G-type, metal-poor stars with HPM, situated in the solar vicinity. These analyses were aimed at determining their Galactic origins. The data utilized for the spectral analyses are detailed in the second section, whereas the third section provides information on the model atmosphere analysis. Following the age calculations using the Bayesian method in the fourth section, kinematic calculations are outlined in the fifth section. The final section evaluates the findings to identify potential scenarios for the Galactic origin of the analyzed stars, and provides concluding remarks.  

\section{Spectroscopic Data}
\label{star selection and spectroscopic data}
The spectra analyzed in this study were obtained from the {\sc ELODIE} spectral library \citep{prugniel2001}, which features a high resolution ($R\sim 42,000$) and high signal-to-noise ratio ($S/N$ $>$ 100). Spectra were obtained using a fiber-fed {\sc ELODIE} echelle spectrometer, covering the wavelength range of 3900 -- 6800 \AA. Table \ref{table:1}lists the key characteristics of HD\,8724 and HD\,195633 and {\sc ELODIE} spectra. To address continuity issues, we used the LIME code \citep{sahin2017} to renormalize the spectra from the spectral library. 

The {\it Gaia} data release 3 \citep[DR3][]{gaiadr32023} radial velocities of HD\,8724 (Gaia DR3 2593188145361730432, $V_{\rm Rad}$=-112.98$\pm$0.13 km s$^{\rm -1}$) and HD\,195633 (Gaia DR3 1748576848010782208, $V_{\rm Rad}$=-45.62$\pm$0.16 km s$^{\rm -1}$) are in agreement with the radial velocities of the stars from their {\sc ELODIE} spectra within error limits (Table \ref{table:1})\footnote{The radial velocities obtained using the cross-correlation method via NARVAL spectrum of the Sun for HD\,8724 and HD\,195633 are -119.84 km s$^{\rm -1}$ and -22.25 km s$^{\rm -1}$, respectively. To convert the radial velocity determined for HD\,8724 to the heliocentric velocity (V$_{\rm Hel}$), a correction of +6.39 km s$^{\rm -1}$ should be made using the observation date information. For HD\,195633, this correction was -23.65 km s$^{\rm -1}$. As a result, the V$_{\rm Hel}$ velocity value we found for HD\,8724 is -113.45 km s$^{\rm -1}$ while the value reported by {\sc ELODIE} is -113.52 km s$^{\rm -1}$. For HD\,195633, the V$_{\rm Hel}$ velocity we found is -45.90 km s$^{\rm -1}$ while the V$_{\rm Hel}$ velocity reported by {\sc ELODIE} is -45.94 km s$^{\rm -1}$. The V$_{\rm Hel}$ velocities reported by {\sc ELODIE} (V$_{\rm Rad}$ in Table \ref{table:1}) agree with the velocity values we found.}. Figure \ref{fig:spec}shows a sample image of the spectra of the analyzed stars.

%Figure 2
\begin{figure*}
    \centering
    \includegraphics[width=8.80cm,height=7.8cm,angle=0]{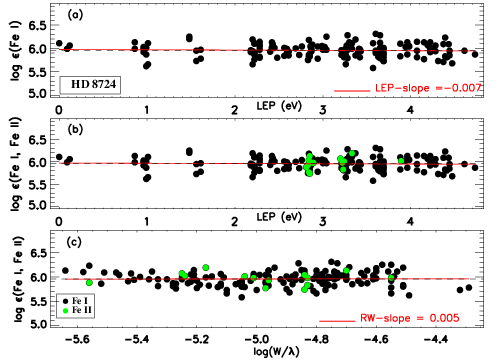}
    \includegraphics[width=8.80cm,height=7.8cm,angle=0]{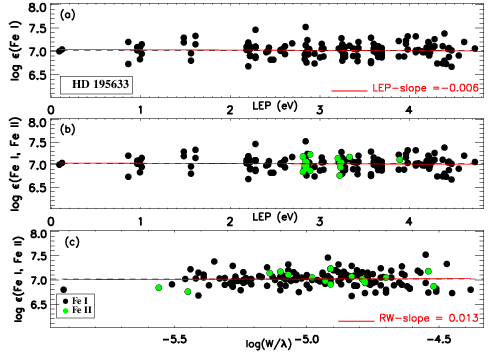}
    \caption{Example image illustrating the calculation of model atmosphere  parameters for HD\,8724 and HD\,195633 as a function of the lower-level excitation potential and reduced equivalent width ($W/\lambda$) using the MOOG stellar atmosphere code. The solid red lines in all panels represent the least-squares fit of the data.}
    \label{fig:moog}
\end{figure*} 

The {\sc LIME} code facilitated line identification in the continuum-normalized spectrum, providing associated atomic data such as the Rowland multiplet number (RMT), $\log~gf$, and lower level excitation energy (L.E.P.). Atomic data were sourced from NIST\footnote{NIST Atomic Spectra Database http://physics.nist.gov/PhysRefData/ASD} and VALD\footnote{VALD Atomic Spectra Database http://vald.astro.uu.se} databases. Most of the identified lines were well-isolated and suitable for equivalent width (EW) analysis, conducted using the {\sc LIME} code.

\section{Model Atmosphere Parameters and Abundances}
The model atmospheres for the two selected stars were calculated using the local thermodynamic equilibrium (LTE; ODFNEW) approach with the ATLAS9 \citep{castelli2004} code. To determine elemental abundances, the line analysis code MOOG \citep{sneden1973}\footnote{Source code of the MOOG software is available at \href{http://www.as.utexas.edu/$\sim$ chris/moog.html}{http://www.as.utexas.edu/$\sim$ chris/moog.html}} operating under LTE conditions was used. For a detailed description of the abundance analysis procedure, see \cite{sahin2009}, \cite{sahin2011}, \cite{sahin2020}, and \cite{sahin2023}. The neutral (Fe\,{\sc i}) and ionized (Fe\,{\sc ii}) iron lines were used to determine model atmosphere parameters -- effective temperature, surface gravity, microturbulence, and metallicity -- of the stars. 

\begin{table*}
\setlength{\tabcolsep}{7.1pt}
\caption{Model atmosphere parameters calculated for HD\,8724, HD\,195633, and the Sun.}\label{modelparametreler}
\label{table:model_param}
\centering
\begin{tabular}{l|c|c|c|c}
\hline
Star	&	$T_{\rm eff}$	& $\log g$ 	&	[Fe/H] &  $\xi$             \\
        &       (K)         & (cgs)     &   (dex)  & (km s$^{\rm -1}$) \\
    \hline
HD\,8724  &	4700$\pm$115	&	1.65$\pm$0.32	&	-1.59$\pm$0.04	&	1.58$\pm$0.50 \\    
HD\,195633 &	6100$\pm$205	&	3.95$\pm$0.35	&	-0.52$\pm$0.05	&	1.26$\pm$0.50 \\	
Sun     &	5790$\pm$125	&	4.40$\pm$0.19	&	 0.00$\pm$0.10	&	0.68$\pm$0.50 \\	
\hline
\end{tabular}
\end{table*}

%Figure 3
\begin{figure*}
    \centering
    \includegraphics[width=0.94\textwidth]{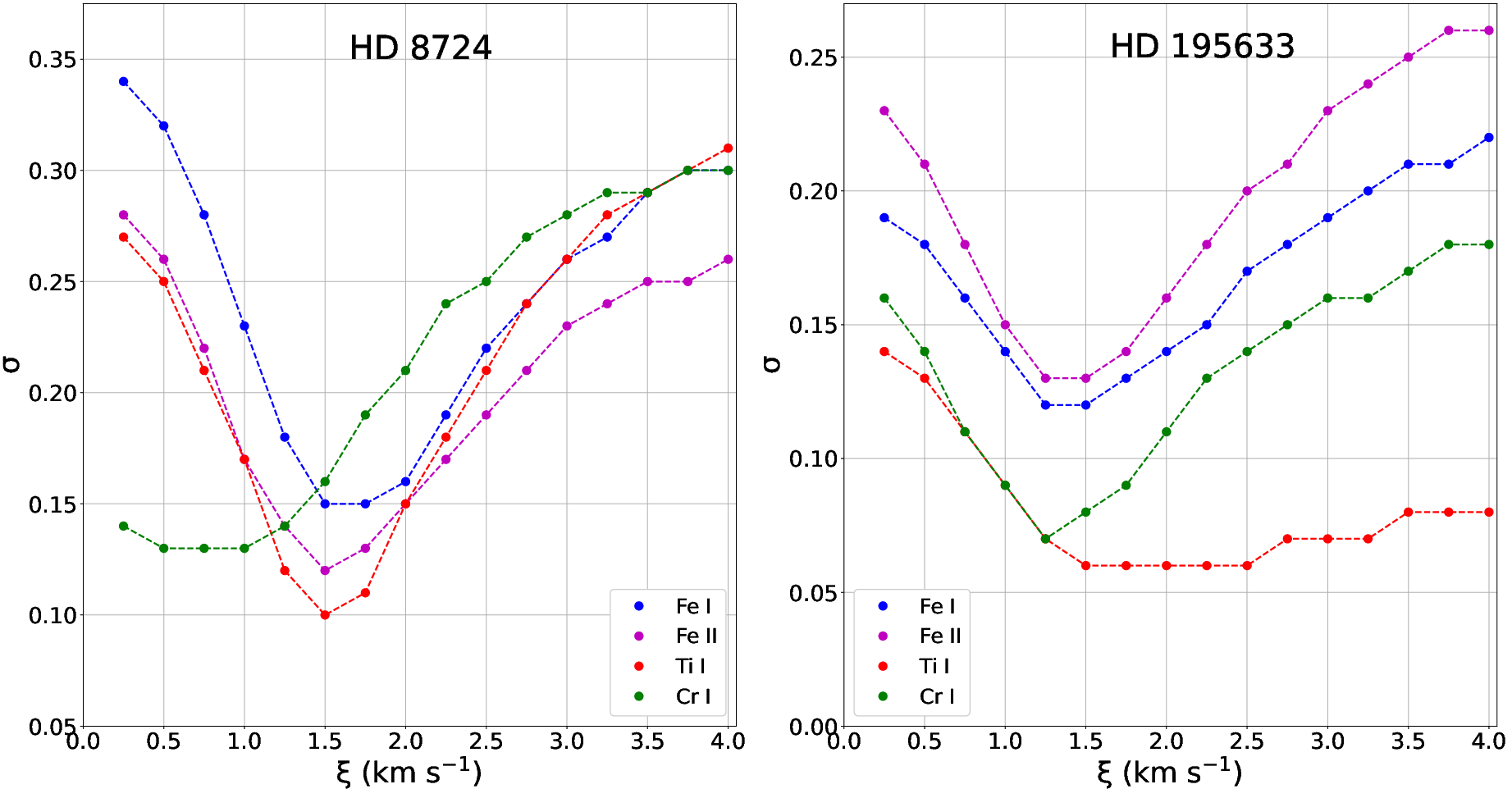}
    \caption{The standard deviation of the Fe, Ti, and Cr abundances from the suite of Fe\,{\sc i}, Fe\,{\sc ii}, Ti\,{\sc i}, and Cr\,{\sc i} lines as a function of $\xi$.}
    \label{fig:dispersion}
\end{figure*}

\begin{table*}
\normalsize
\setlength{\tabcolsep}{5.6pt}
\caption{The model atmosphere parameters for HD\,8724 and HD\,195633 were compiled from the 2020 version of the PASTEL catalog \citep{soubiran2016}. The results of the afresh determination of the model parameters in this study are also included.}
\label{table:lit}
\centering
\begin{tabular}{l|c|c|c||l|c|c|c}

\hline
\multicolumn{4}{c}{HD\,8724} & \multicolumn{4}{c}{HD\,195633} \\
\hline
$T_{\rm eff}$ & $\log g$  &  [Fe/H]	&   Reference  & $T_{\rm eff}$ & $\log g$  & [Fe/H]	&   Reference\\
(K)           & (cgs)     &  (dex)  &              &  (K)          &   (cgs)   & (dex)  &            \\
\hline
4700$\pm$115   & 1.65$\pm$0.32 &-1.59$\pm$0.04& This study     & 6100$\pm$205 & 3.95 $\pm$ 0.35& -0.52$\pm$0.05 & This study	   \\
 4679 		  & --            & --      	      &  01  &5960	              & 4.07		        & -0.91		          & 15     \\
 4560	      & 1.29 	      &-1.76$\pm$0.04     &  02  &6024	              & 3.99		        & -0.59		          & 16 	      \\
 4790$\pm$30	  & 1.66$\pm$0.12 &-1.55$\pm$0.07     & 03   &6117$\pm$49        & 3.98$\pm$0.08	    & -0.58$\pm$0.03	  & 17 	  \\
 4632$\pm$50   & --	    	  &--	  	          & 04   &6154$\pm$37        & 4.25$\pm$0.05	    & -0.51$\pm$0.03	  & 18 	  \\
 4577 		  & 1.49          & -1.69 	          & 05   &6131$\pm$111       & --		            &	  --		      & 19  \\
 4539$\pm$51	  & --	          &--	  	          & 06   &5990	              & 3.79		        & -0.62		          & 20 	  \\
 4600$\pm$100  & 1.50$\pm$0.30 &-1.65$\pm$0.15     & 07   &6063$\pm$54        & --		            &	  --		      & 21 	  \\ 
 4625$\pm$40	  & 1.20$\pm$0.60 &-1.83$\pm$0.04     & 08   &5936	              & 3.76		        & -0.73		          & 22     \\
 4500 		  & 1.20 	      & -1.84 	          & 09   &5987$\pm$50        & 3.95$\pm$0.15	    & -0.54$\pm$0.08	  & 23 	  \\
 4500$\pm$100  & 1.02$\pm$0.20 &-1.92$\pm$0.16     & 10   &5902	              & 3.83		        & -0.50		          & 24 	      \\
 4535$\pm$45	  & --	   	      &--	  	          & 11   &6000$\pm$100       & 3.80$\pm$0.30	    & -0.55$\pm$0.15	  & 7   \\
 4500 		  & 1.20 	      &-1.84 		      & 12   &6000$\pm$40        & 3.90$\pm$0.06	    & -0.65$\pm$0.04	  & 8   \\
 4800 		  & 1.90          & -1.52 	          & 13   &5793	              & 3.80		        & -0.90		          & 25 	  \\
 4624 		  & 1.50          & -1.60 	          & 14   &6146	              & 3.70		        & -1.10		          & 26 	  \\
 \hline
\end{tabular}
\begin{minipage}{17cm}
\vskip 5pt
[01] \cite{gaiadr22018}, [02] \cite{mashonkina2017}, [03] \cite{matrozis2013}, [04] \cite{hernandez2009}, [05] \cite{melendez2008}, [06] \cite{ramirez2005}, [07] \cite{mishenina2001}, [08] \cite{fulbright2000}, [09] \cite{burris2000}, [10] \cite{tomkin1999}, [11] \cite{alonso1999}, [12] \cite{pilachowski1996}, [13] \cite{shetrone1996}, [14] \cite{gratton1984}, [15] \cite{gaiadr32023}, [16] \cite{luck2016}, [17] \cite{bensby2014}, [18] \cite{sousa2011}, [19] \cite{casagrande2011}, [20] \cite{zhang2006}, [21] \cite{masana2006}, [22] \cite{gratton2003}, [23] \cite{nissen2002}, [24] \cite{qiu2002}, [25] \cite{rebolo1988}, [26] \cite{sneden1974}
\end{minipage}
\end{table*}

%Table 4
\begin{table*}
\setlength{\tabcolsep}{1.0pt}
\caption{The element abundances [$X$/Fe] calculated for stars HD\,8724 and HD\,195633 are shown. Simultaneously, the Solar abundances obtained in this study (TS) and those reported by \citet{asplund2009}[ASP] are provided. Abundances in bold are those calculated via spectrum synthesis method.}
\label{table:abund}
\centering
\begin{tabular}{l|cccc|ccccc|ccc|c}
\hline
     	& \multicolumn{4}{c|}{HD\,8724} & \multicolumn{5}{c|}{HD\,195633} & \multicolumn{3}{c|}{Sun} &  \\
\cline{2-4}
\cline{5-8}
\cline{9-13}
Species   &  [$X$/Fe]$^{\rm \ast}$ & $\sigma_{\rm abs}$ &$n$ &[$X$/Fe]$^{\rm \dag}$ &   [$X$/Fe]$^{\rm \ast}$ $\sigma_{\rm abs}$ &$n$ & [$X$/Fe]$^{\rm \dag}$ &[$X$/Fe]$^{\rm \ddag}$ & $\log\epsilon_{\rm \odot}$(X$_{\rm TS}$) & $n$ &$\log\epsilon_{\rm \odot}$(X$_{\rm ASP}$)&$\Delta\log\epsilon_{\rm \odot}$(X) \\
\cline{2-4}
\cline{5-8}
\cline{9-13} 
  &  (dex) &  &   & (dex) &  (dex) & & & (dex)  & (dex) & (dex)   &     & (dex) & (dex) \\
 \hline
\textbf{Mg\,{\sc i}}	  &   \textbf{0.35$\pm$0.05} & \textbf{0.14} &	1   & 0.22  & \textbf{0.07$\pm$0.05} & \textbf{0.08}   & 1   & -0.03& 0.20   & \textbf{7.58$\pm$0.00}  &  1  & 7.60$\pm$0.04 & -0.02 \\
Si\,{\sc i}	  &   0.43$\pm$0.08 & 0.05 &	4   & 0.38  & 0.16$\pm$0.08 & 0.06  & 8  & 0.13 & 0.13    & 7.48$\pm$0.04  & 11   & 7.51$\pm$0.03 & -0.03 \\
Ca\,{\sc i}	  &   0.26$\pm$0.06 & 0.17 &	7   & 0.37  & 0.11$\pm$0.06 & 0.17  & 7  & 0.21 & 0.19    & 6.35$\pm$0.03  &  8   & 6.34$\pm$0.04 & 0.01 \\
Sc\,{\sc ii}  &  -0.07$\pm$0.07 & 0.16 &	4   & ---   & 0.08$\pm$0.07 & 0.18  & 3   & ---  & ---    & 3.30$\pm$0.06  &  4   & 3.15$\pm$0.04 & 0.15 \\
Ti\,{\sc i}	  &   0.15$\pm$0.06 & 0.26 &	16  & ---   & 0.06$\pm$0.06 & 0.18  & 8  & ---  & 0.11    & 4.97$\pm$0.02  & 28   & 4.95$\pm$0.05 & 0.02 \\
Ti\,{\sc ii}  &   0.29$\pm$0.07 & 0.18 &	5   & ---   & 0.14$\pm$0.06 & 0.17  & 6  & ---  & ---     & 5.00$\pm$0.03  & 10   & 4.95$\pm$0.05 & 0.05 \\
V\,{\sc i}    &  -0.11$\pm$0.07 & 0.17 &	1   & ---   & ---           & ---      &    & ---  & ---     &3.99$\pm$0.05   &  2   & 3.93$\pm$ 0.08 & 0.06 \\
Cr\,{\sc i}	  &  -0.14$\pm$0.08 & 0.27 &	8   & ---   & -0.05$\pm$0.06& 0.19  & 10  & ---  & -0.01  &5.71$\pm$0.02   & 14   & 5.64$\pm$0.04 & 0.07 \\
Cr\,{\sc ii}  &   0.05$\pm$0.11 & 0.13 &	3   & ---   & -0.02$\pm$0.11& 0.17  & 3   & ---  & ---    &5.64$\pm$0.08   &  3   & 5.64$\pm$0.04 & 0.00 \\
Mn\,{\sc i}	  &  -0.62$\pm$0.08 & 0.17 &  5   & ---   & -0.23$\pm$0.09  & 0.15  & 4   & ---  & -0.12  &5.64$\pm$0.04   &  7   & 5.43$\pm$0.05 & 0.21 \\
Fe\,{\sc i}	  &  -0.03$\pm$0.05 & 0.19 &  68  & ---   &  0.00$\pm$0.05  & 0.18  & 67 & ---  & ---     &7.54$\pm$0.01   & 84   & 7.50$\pm$0.04 & 0.04 \\
Fe\,{\sc ii}  &   0.00$\pm$0.07 & 0.17 &  12  & ---   &  0.00$\pm$0.07  & 0.19  & 13  & ---  & ---    &7.54$\pm$0.03   & 13   & 7.50$\pm$0.04 & 0.04 \\
Co\,{\sc i}	  &  -0.10$\pm$0.10 & 0.11 &  3   & ---   & -0.04$\pm$0.10  & 0.11  & 2   & ---  & ---    & 4.95$\pm$0.07  &  6   & 4.99$\pm$0.07 & -0.04 \\
Ni\,{\sc i}	  &  -0.09$\pm$0.06 & 0.14 &	12  & ---   & -0.01$\pm$0.07& 0.15  & 15  & ---  & 0.00   &6.26$\pm$0.03   & 24   & 6.22$\pm$0.04 & 0.04 \\
\textbf{Zn\,{\sc i}}   &   \textbf{-0.03$\pm$0.05} & \textbf{0.12} &	1   &  -0.09$^{\rm a}$     & \textbf{-0.09$\pm$0.08}  & \textbf{0.14} & 2  & -0.05$^{\rm a}$  & ---    &\textbf{4.62$\pm$0.02}   &  2   & 4.56$\pm$0.05 & 0.06 \\
\textbf{Y\,{\sc ii}}   &  \textbf{-0.11$\pm$0.05}  & \textbf{0.22} &	2  & -0.16 & \textbf{-0.18$\pm$0.07}& \textbf{0.11} & 2   & 0.09 & ---    &\textbf{2.21$\pm$0.02}   &  2   & 2.21$\pm$0.05 & 0.00 \\
\textbf{Zr\,{\sc ii}}  &   \textbf{0.22$\pm$0.05}  & \textbf{0.31} &	1  & ---   & \textbf{-0.09$\pm$0.05}& \textbf{0.13}  & 1   &   ---   & ---    &\textbf{2.63$\pm$0.00}   & 1   & 2.58$\pm$0.04 & 0.05 \\
\textbf{Ba\,{\sc ii}}  &   \textbf{0.11$\pm$0.14}  & \textbf{0.35} &	2   & 0.18  & \textbf{-0.05$\pm$0.10}& \textbf{0.25} & 2   & 0.12 & ---    &\textbf{2.34$\pm$0.01}   & 2   & 2.18$\pm$0.09 & 0.16 \\
\textbf{Ce\,{\sc ii}}  &   \textbf{0.06$\pm$0.05}  & \textbf{0.15} &	1   & -0.12 & ---              & ---  & ---  & 0.11 & ---    & \textbf{1.57$\pm$0.00}  &  1   & 1.58$\pm$0.04 & -0.01 \\
\textbf{Nd\,{\sc ii}}  &   \textbf{0.21$\pm$0.05}  & \textbf{0.10} &	1   & 0.15  &  ---             & --- & ---   & 0.10 & ---    &\textbf{1.34$\pm$0.00}   &  1   & 1.42$\pm$0.04 & -0.08 \\
\textbf{Sm\,{\sc ii}}  &   \textbf{0.24$\pm$0.05}  & \textbf{0.16} &	1   & ---   &  \textbf{0.34$\pm$0.05} & \textbf{0.20}& 1 & ---  & ---    & \textbf{0.97$\pm$0.00}  &  1   & 0.96$\pm$0.04 & 0.01 \\
\hline
\end{tabular}
\begin{tablenotes}
      \item ($\ast$): This study (TS); ($\dag$): \citet{mishenina2001}; ($\ddag$): \citet{zhang2006}; ($^{\rm a}$): The [Zn\,{\sc i}/Fe] ratios are from \citet{mishenina2002} with the same model atmosphere parameters as in \citet{mishenina2001}.\\
$\Delta \log \epsilon_{\odot}(X)=\log \epsilon_{\odot}(X_{\rm TS}) - \log \epsilon_{\odot}(X_{\rm ASP})$
    \end{tablenotes}
\end{table*}

Effective temperature ($T_{\rm eff}$) determination employed the excitation-balance method (sensitive to neutral spectral lines with a broad range of excitation potentials) for Fe\,{\sc i}. The value of $T_{\rm eff}$ was selected such that the abundance was independent of the L.E.P. of a line. The microturbulence parameter ($\xi$), representing small-scale gas movement in the stellar atmosphere, was determined by the usual requirement that the abundance of neutral or single-ionized atoms (i.e., Fe\,{\sc i}) be independent of the reduced EW ($W/\lambda$), here assuming LTE. For our sample of Fe\,{\sc i} lines, these two conditions are imposed simultaneously (upper and middle panels in Figure \ref{fig:moog}). The microturbulence may also be determined separately from the Fe\,{\sc i}, Ti\,{\sc i} and Cr\,{\sc i}. For a given model, we computed the dispersion in the Fe, Ti, and Cr abundances over a range in $\xi$ from $\approx$0.3 to 4.0 km s$^{\rm -1}$. Figure \ref{fig:dispersion} shows the dispersion $\sigma$ for the Fe\,{\sc i}, Fe\,{\sc ii}, Ti\,{\sc i}, and Cr\,{\sc i}. The dispersion test applied to HD\,8724 (left panel in the Figure \ref{fig:dispersion}) using Fe\,{\sc i} lines yielded a microturbulent velocity ranging from 1.5 to 1.8 km s$^{\rm -1}$. This method also provided a value of 1.5 km s$^{\rm -1}$ for both Fe\,{\sc ii} and Ti\,{\sc i} lines. However, the dispersion test further suggested an upper limit of 1.0 km s$^{\rm -1}$ for Cr\,{\sc i} lines. Similarly, for HD\,195633, the analysis using Fe\,{\sc i} and Fe\,{\sc ii} lines indicated a microturbulent velocity range of $\approx$1.3 to 1.5 km s$^{\rm -1}$. Notably, the dependence of the microturbulent velocity on the neutral titanium lines weakens beyond 1.5 km s$^{\rm -1}$. When both methods are evaluated together, the measurement uncertainty for both stars was estimated to be 0.5 km s$^{\rm -1}$.

The determination of the surface gravity parameter ($\log~g$) involved analyzing abundances calculated with MOOG for Fe, ensuring the presence of ionization equilibrium (where Fe\,{\sc i} and Fe\,{\sc ii} lines provide the same iron abundance). Finally, the metallicity ([Fe/H]) was refined through an iterative process to achieve convergence between the derived abundance of iron and the abundance initially adopted for model atmosphere construction. Convergence was achieved by adjusting the effective temperature ($T_{\rm eff}$), surface gravity ($\log g$), and microturbulent velocity ($\xi$) of the model. Figure \ref{fig:moog} illustrates a summary of the relationship between these physical parameters used to determine the stellar model parameters using the classical spectroscopic method (i.e., the ionisation and excitation equilibria of Fe lines) for HD\,8724 (left panel) and HD\,195633 (right panel). 

The uncertainty in the derived surface temperature is due to the error in the slope of the relationship between Fe\,{\sc i} abundance and L.E.Ps of the lines. For HD\,8724, a noticeable change in the slope was observed for a temperature variation of $\pm$115 K in the assumed model (see top panel of Figure \ref{fig:moog}). Similarly, a 1$\sigma$ difference in the abundance [X/H] between the neutral and ionised lines of Fe corresponds to a $\approx$0.3 dex change in $\log g$. For HD\,195633, the same method indicates errors of $\pm$205 K in $T_{\rm eff}$ and 0.35 dex in $\log g$.  The atmospheric model parameters for the stars are listed in Table \ref{table:model_param}. 

Examination of model atmospheric parameters reported in the literature for these stars revealed considerable variations.Table \ref{table:lit} provides earlier determinations of the stellar parameters from the PASTEL catalog \citep{soubiran2016}. Figure \ref{fig:A1} displays graphical representations of the data presented in Table \ref{table:lit}, depicting $T_{\rm eff}$, $\log~g$, and [Fe/H] reported in the literature, and the model parameters obtained in this study for the two stars along with their uncertainties. A comparison of the atmospheric model parameters reported in this study with those reported by \cite{mishenina2001} for HD\,8724 shows differences in $T_{\rm eff}$, $\log~g$ and [Fe/H] are -100 K, -0.15 cgs, and -0.11 dex, respectively. HD\,8724 is listed in the {\it Gaia} DR3 \citep{gaiadr32023} catalog. The spectroscopic temperature of the star obtained in this study corresponds to the temperature value obtained from the {\it Gaia} consortium (Table \ref{table:lit}). For HD\,195633, the differences in the model atmospheric parameters for $T_{\rm eff}$, $\log~g$ and [Fe/H] were similar to those in \cite{mishenina2001} for HD\,8724. 

%Table 5
\begin{table*}
\setlength{\tabcolsep}{2.8pt}
\caption{Sensitivity of the derived abundances to the uncertainties of $\Delta T_{\rm eff} = +$115 K, $\Delta \log g = +$0.32 cgs, and $\Delta \xi = \pm$0.5 km s$^{\rm -1}$ in the model atmosphere
parameters of HD\,8724 for $T_{\rm eff} =$ 4700 K, $\log g$ = 1.65 cgs, and $\xi$ = 1.58 km s$^{\rm -1}$. The derived uncertainties of $\Delta T_{\rm eff} = +$205 K, $\Delta \log g = +$0.35 cgs, and $\Delta \xi = \pm$0.5 km s$^{\rm -1}$ in the model parameters of HD\,195633 for $T_{\rm eff} =$ 6100 K, $\log g$ = 3.95 cgs, and $\xi$ = 1.26 km s$^{\rm -1}$ were also reported. The uncertainties in bold are
those calculated via spectrum synthesis method.}
\label{table:abund_err}
\centering
\begin{tabular}{l|c|c|c|c||c|c|c|c}
\hline
& \multicolumn{4}{c||}{$\Delta \log \epsilon$ (HD\,8724)} & \multicolumn{4}{c}{$\Delta \log \epsilon$ (HD\,195633)} \\
\cline{2-9}
Species	&$\Delta T_{\rm eff}$& $\Delta \log g$ 	&$\Delta \xi$&  $\Delta \xi$  &	$\Delta T_{\rm eff}$& $\Delta \log g$ 	&$\Delta \xi$&  $\Delta \xi$ \\
        &       (+115 K)         & (+0.32 cgs)     &  (+0.50 km s$^{\rm -1}$)  & (-0.50 km s$^{\rm -1}$) &       (+205 K)         &(+0.35 cgs)     &   (+0.50 km s$^{\rm -1}$)  & (-0.50 km s$^{\rm -1}$)\\
\cline{2-9}
    \hline
\textbf{Mg\,{\sc i}}	&	\textbf{0.13}	&	\textbf{0.02}		&	\textbf{-0.06}	&	\textbf{0.08}&	\textbf{0.06}	&	\textbf{-0.04}		&	\textbf{-0.03}	&	\textbf{0.02}\\
Si\,{\sc i}	&	0.04	&	0.02		&	-0.02	&	0.02&	0.06	&	0.01		&	-0.02	&	0.02\\
Ca\,{\sc i}	&	0.11	&	 -0.03		&	-0.13	&	0.17&	0.12	&	-0.05		&	-0.11	&	0.11\\
Sc\,{\sc ii}&	 0.01	&	0.11		&	-0.11	&	0.16&	0.04	&	0.13		&	-0.12	&	0.14\\
Ti\,{\sc i}	&	0.20	&	 -0.03		&	-0.17	&	0.23&	0.17	&	0.00		&	-0.06	&	0.09\\
Ti\,{\sc ii}&	 0.01	&	0.11		&	-0.14	&	0.20&	0.04	&	0.11		&	-0.12	&	0.12\\
V\,{\sc i}  &	0.17	&	 -0.01		&	-0.02	&	0.03&	 --	    &	 --	        &	 --	    &	  --	\\
Cr\,{\sc i}	&	0.18	&	 -0.03		&	-0.20	&	0.31&	0.15	&	-0.02		&	-0.12	&	0.14\\
Cr\,{\sc ii}&	-0.04	&	 0.10		&	-0.07	&	0.10&	-0.01	&	0.14		&	-0.10	&	0.14\\
Mn\,{\sc i}	&	0.15	&	 -0.02		&	-0.07	&	0.11&	0.13	&	-0.01		&	-0.08	&	0.12\\
Fe\,{\sc i}	&	0.14	&	 -0.02		&	-0.13	&	0.18&	0.15	&	-0.02		&	-0.10	&	0.12\\
Fe\,{\sc ii}&	-0.03	&	0.12		&	-0.12	&	0.19&	0.01	&	0.13		&	-0.14	&	0.18\\
Co\,{\sc i} &	0.11	&	 0.00		&	-0.01	&	0.02&	0.11	&	0.00		&	-0.02	&	0.02\\
Ni\,{\sc i}	&	0.12	&	 0.00		&	-0.07	&	0.12&	0.14	&	0.00		&	-0.05	&	0.07\\
\textbf{Zn\,{\sc i}} &	\textbf{-0.03}	&	\textbf{0.03}		&	\textbf{-0.11}	&	\textbf{0.14}&	\textbf{0.05}	&	\textbf{0.00}		&	\textbf{-0.13}	&	\textbf{0.15}\\
\textbf{Y\,{\sc ii}} &	\textbf{0.02}	&	\textbf{0.09}		&	\textbf{-0.20}	&	\textbf{0.25}&	\textbf{-0.02}	&	\textbf{0.06}		&	\textbf{-0.09}	&	\textbf{0.15}\\
\textbf{Zr\,{\sc ii}}&	\textbf{-0.06}	&	\textbf{0.00}		&	\textbf{-0.30}	&	\textbf{0.30}&	\textbf{-0.02}	&	\textbf{0.05}		&	\textbf{-0.12}	&	\textbf{0.08}\\
\textbf{Ba\,{\sc ii}}&	\textbf{0.07}	&	\textbf{0.11}		&	\textbf{-0.33}	&	\textbf{0.46}&	\textbf{0.09}	&	\textbf{0.12}		&	\textbf{-0.21}	&	\textbf{0.34}\\
\textbf{Ce\,{\sc ii}}&  \textbf{0.05}	&	\textbf{0.12}		&	\textbf{-0.07}	&	\textbf{0.09}&	--	&	--	&	--	&	--	\\
\textbf{Nd\,{\sc ii}}&	\textbf{0.02}	&	\textbf{0.09}		&	\textbf{-0.05}	&	\textbf{0.04}&	--	&	--	&	--	&	--	\\
\textbf{Sm\,{\sc ii}}&	\textbf{0.07}	&	\textbf{0.14}		&	\textbf{-0.02}	&	\textbf{0.04}&	\textbf{0.11}	&	\textbf{0.16}		&	\textbf{0.03}	&	\textbf{0.05}\\
\hline
\end{tabular}
\end{table*}

Uncertainties in atomic data ($\log gf$) were assessed by deriving solar abundances from stellar spectral lines. It is important to recognize that 1D LTE models used for solar abundances have limitations owing to 3D and NLTE effects, and these effects might be more pronounced for stars such as HD\,8724. A high-resolution solar spectrum from \cite{kurucz1984} was used for this analysis, employing established model parameters for the solar photosphere (listed in Table \ref{table:model_param}). The resulting solar abundances (presented in Table \ref{table:abund}) served as reference points for determining the abundance ratios of various elements relative to iron ([element/Fe]) in this study. Therefore, in this study, we employed a differential abundance analysis approach. This method offers several advantages. Firstly, it minimizes the impact of uncertainties associated with $\log gf$ values. Secondly, it reduces the influence of potential inaccuracies arising from the LTE assumption, which can affect absolute abundance determinations. Finally, differential abundance analysis effectively addresses uncertainties related to EW measurements.
Building on the previous analysis where Fe\,{\sc i} and Fe\,{\sc ii} abundances were employed to constrain stellar model parameters, we now address potential non-LTE (non-Local Thermodynamic Equilibrium) effects on iron. Fortunately, \citet{bergemann2012, lind2012, bensby2014} suggest these effects are negligible for Fe\,{\sc ii} lines. To account for non-LTE effects on the Fe lines, we adopted the 1D non-LTE investigation by \citet{lind2012} utilizing the INSPECT program v1.0 (see \citet{lind2012}). This program provides non-LTE corrections for various elements. For the commonly used Fe lines in the analysis of HD\,8724, the INSPECT program yielded non-LTE corrections\footnote{$\Delta$$\log\epsilon$(Fe\,{\sc i}) = $\log\epsilon$(Fe\,{\sc i})$_{\rm NLTE}$ - $\log \epsilon$(Fe\,{\sc i})$_{\rm LTE}$} of 0.051$\pm$0.027 dex for Fe\,{\sc i} and a negligible 0.000$\pm$0.002 dex for Fe\,{\sc ii}. 

The impact of non-LTE effects on other elements varied. While Mg\,{\sc i} and Si\,{\sc i} exhibited minimal influence, Ca\,{\sc i} and Cr\,{\sc i} showed corrections ranging from 0.02 dex to 0.17 dex. The most significant correction was observed for Ti\,{\sc i}, exceeding 0.3 dex. However, caution is advised when interpreting the correction for Co\,{\sc i}, which is even larger (0.345$\pm$0.180 dex) but has a high associated uncertainty. 

Similar to HD\,8724, the impact of non-LTE effects on elemental abundances in HD\,195633 varied. Mg\,{\sc i} and Si\,{\sc i} exhibited negligible corrections, with Mg\,{\sc i} showing a small positive adjustment of 0.027 dex and Si\,{\sc i} showing a small negative correction of -0.008 dex. Ca\,{\sc i} displayed a small negative correction (-0.028 dex) but with a relatively high uncertainty (0.148 dex). The correction for Ti\,{\sc i} shows contrasting result, showing a positive correction of 0.208 dex. Cr\,{\sc i} also displayed a positive correction of 0.136 dex. In contrast, other elements with relatively large corrections include Ti\,{\sc i} at +0.208 dex and Cr\,{\sc i} at +0.136 dex. In this analysis, the non-LTE correction for Fe\,{\sc i} was only 0.027 dex, and Fe\,{\sc ii} exhibited a negligible correction close to zero (0.001 dex) with minimal uncertainty. This suggests that under the specific atmospheric model parameters for HD\,195633, non-LTE effects have a negligible impact on Fe\,{\sc i} and Fe\,{\sc ii} abundance calculations. Similar to HD\,8724, Co\,{\sc i} displayed the most significant correction in HD\,195633, with a value of +0.301 dex. However, caution is advised due to the potential for high uncertainties associated with such large corrections. 
%\vfill
%Figure 4
\begin{figure}
    \centering
    \includegraphics[width=0.47\textwidth]{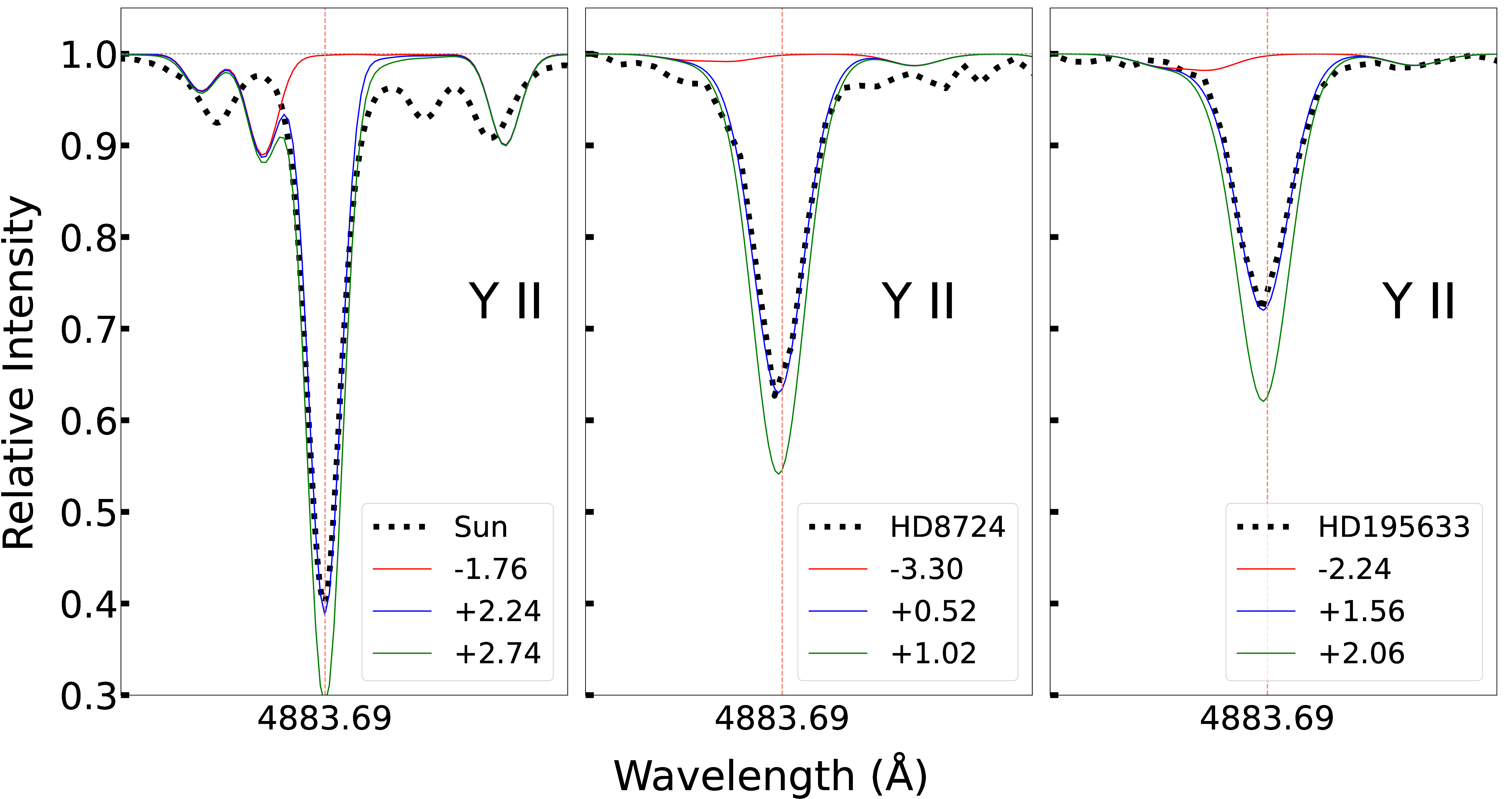}
            \includegraphics[width=0.47\textwidth]{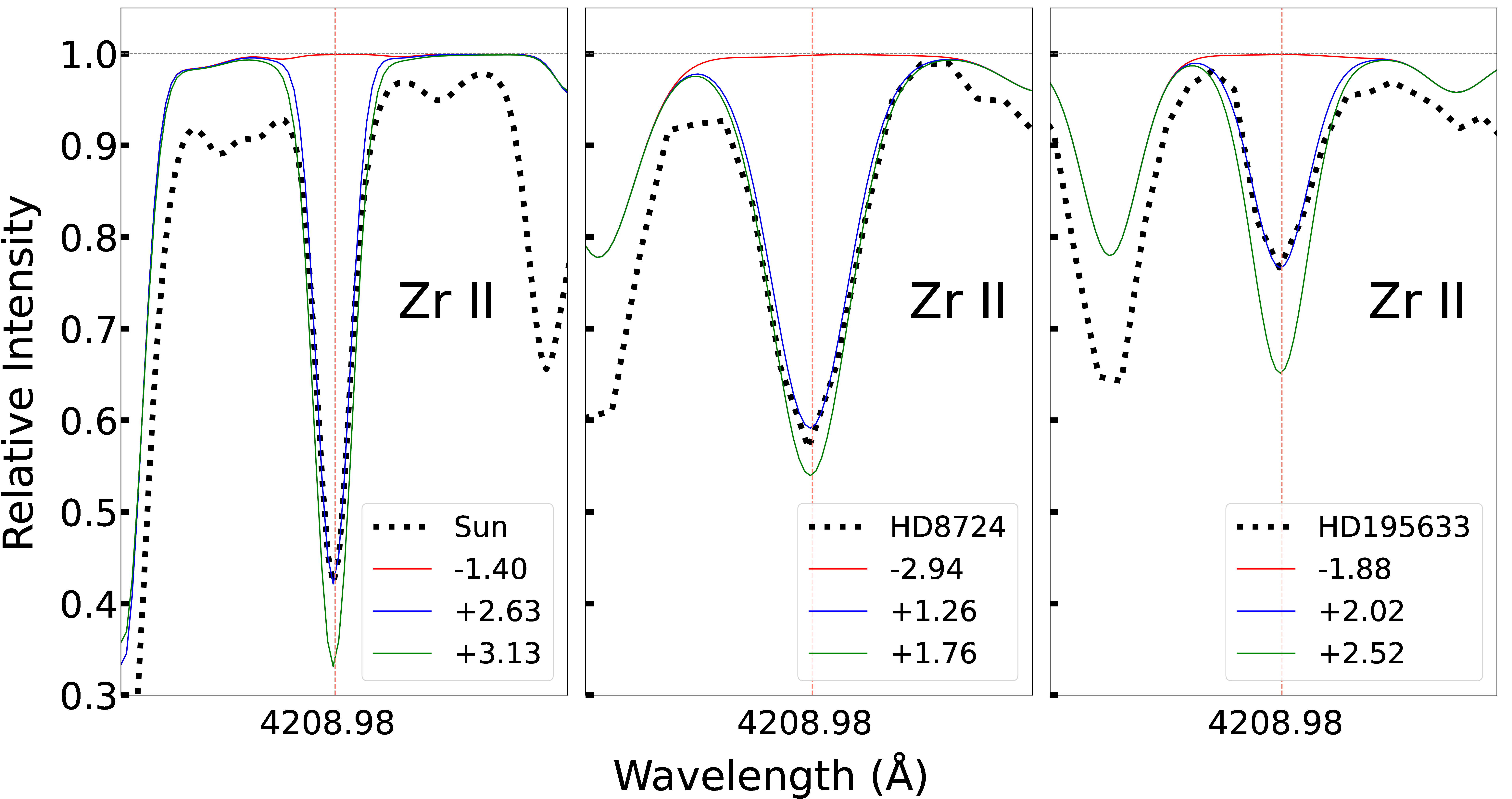}
                       \includegraphics[width=0.47\textwidth]{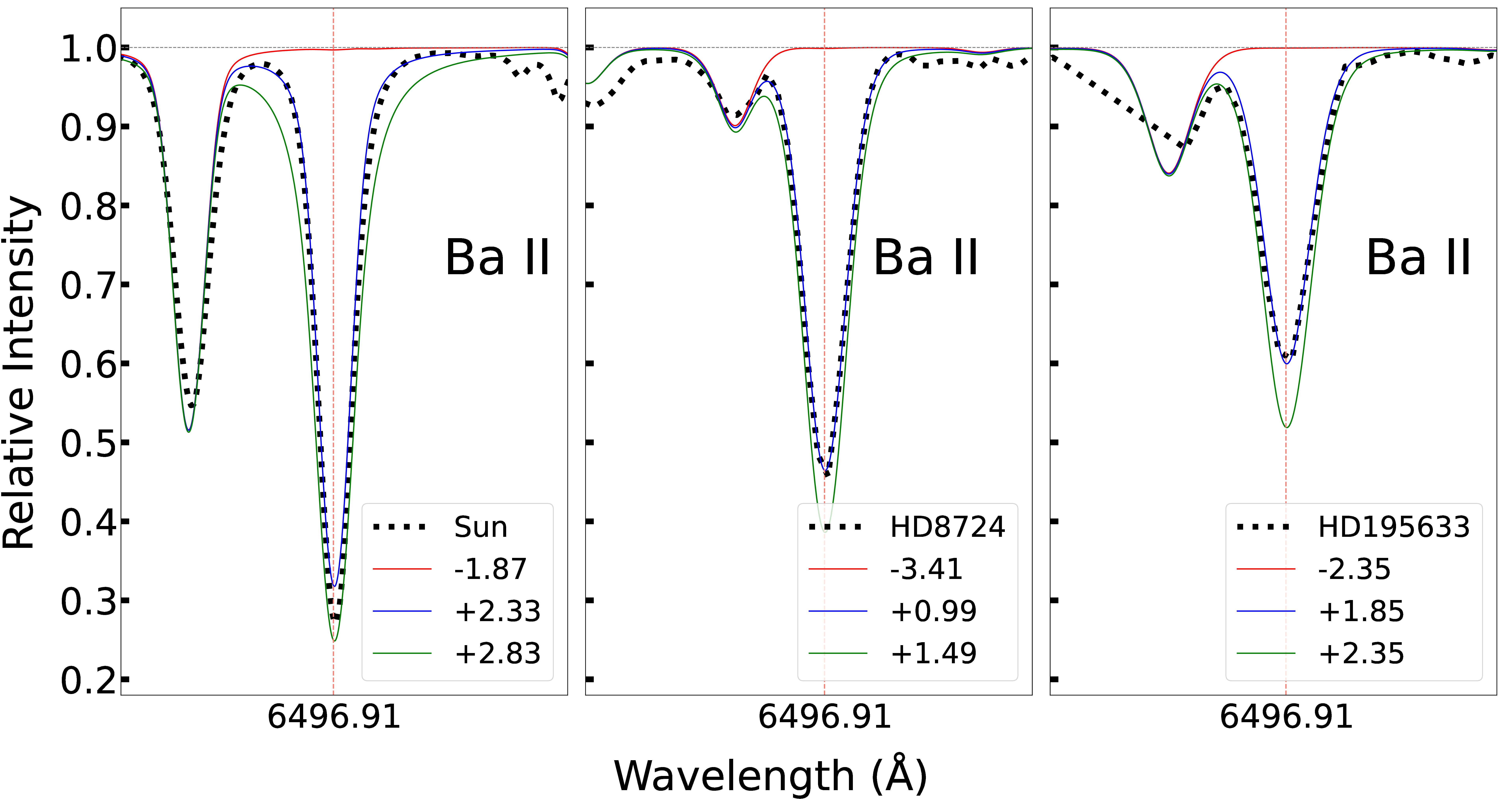}
    \caption{The observed (filled circles) and computed (full blue line) line profiles for some of the ionized metal lines used in the analysis of HD\,8724, HD\,195633 and the Sun. The computed profiles illustrate the synthetic spectra for the three varying logarithmic abundances. The red lines are the spectra computed with no contribution from those ionized metal lines.}
    \label{fig:Y_Zr_Ba}
\end{figure}

%Figure 5
\begin{figure}
    \centering
                        \includegraphics[width=0.47\textwidth]{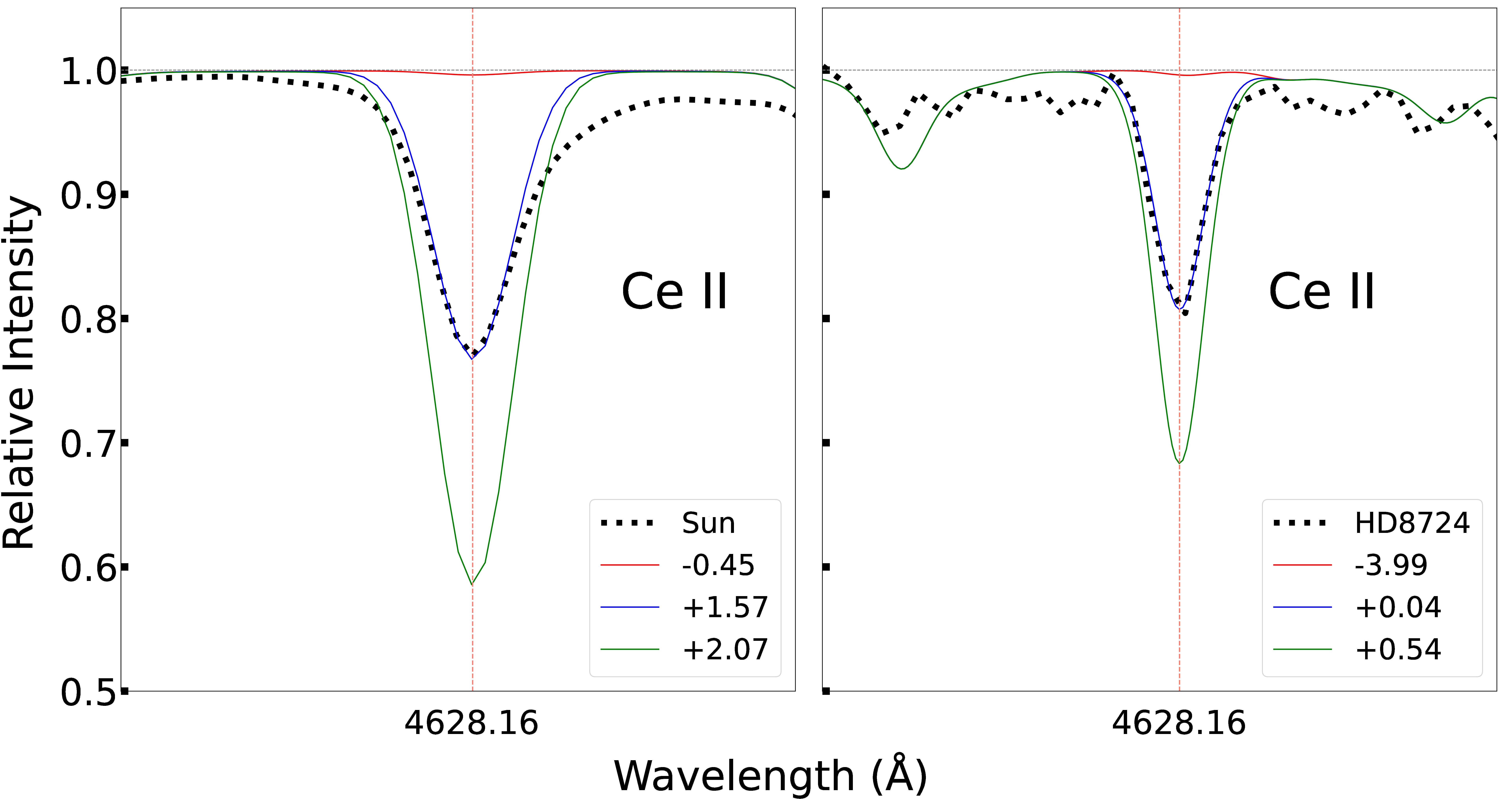}
                        \includegraphics[width=0.47\textwidth]{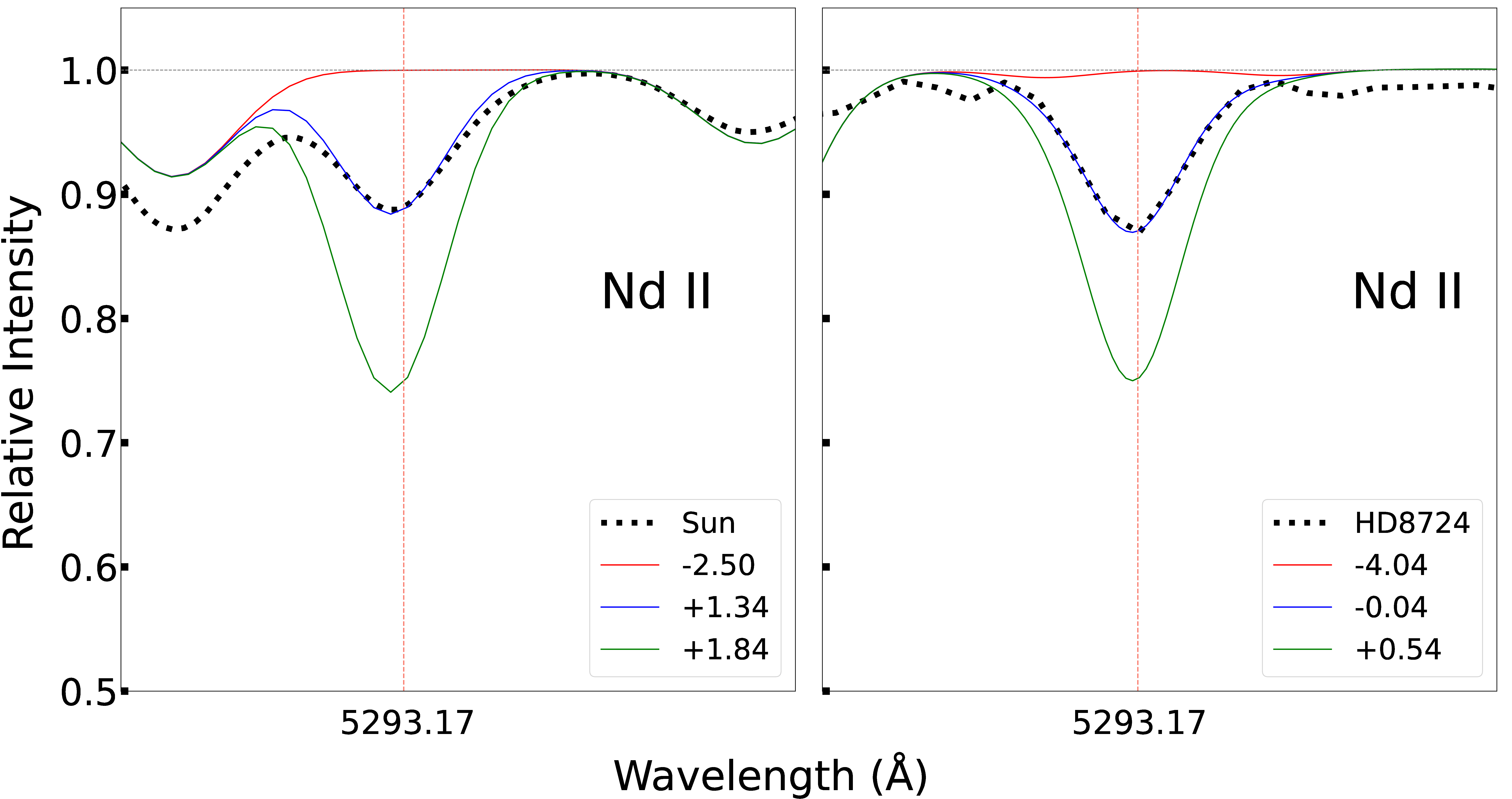}
                        \includegraphics[width=0.47\textwidth]{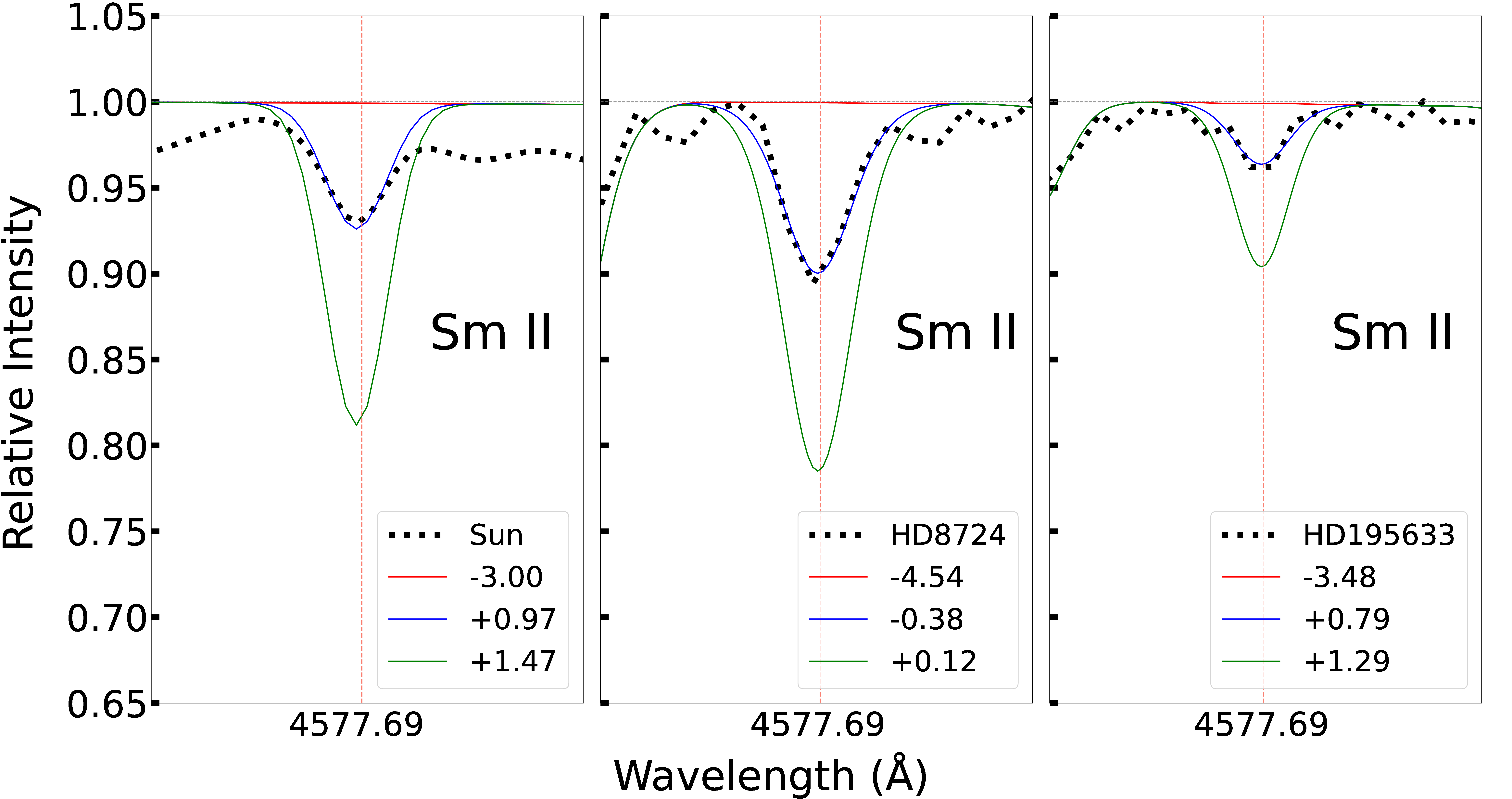}

    \caption{The observed (filled circles) and computed (full blue line) line profiles for some of the ionized metal lines used in the analysis of HD\,8724, HD\,195633 and the Sun. The computed profiles illustrate the synthetic spectra for the three varying logarithmic abundances. The red lines are the spectra computed with no contribution from those ionized metal lines.}
    \label{fig:Ce_Nd_Sm}
\end{figure}

This study reports the abundance of 18 elements (21 species) for HD\,8724 and 15 elements (18 species) for HD\,195633 (Table \ref{table:abund}). In Table \ref{table:abund}, [X/Fe] represents the logarithmic abundance of an element (X) relative to iron (Fe). The uncertainty in [X/Fe] is the square root of the sum of the quadratures of the errors in [X/H] and [Fe/H]. The error in [X/H] is the square root of the sum of the quadratures of the errors (1$\sigma$ line-to-line scatter in abundances) in the stellar and solar logarithmic abundances (see Table \ref{table:abund}). Importantly, the final abundance for each element is the average of the abundances determined from all its spectral lines. The uncertainty in this average abundance is derived from the standard error of the mean. The estimated formal errors for the abundances due to uncertainties in the atmospheric parameters $T_{\rm eff}$, $\log~g$ and $\xi$ are summarized in Table \ref{table:abund_err} for changes with respect to the model of +115 K, +0.32 cgs, and $\pm$0.5 km s$^{\rm -1}$ for HD\,8724. They are with respect to the model of +205 K, +0.35 cgs units and $\pm$ 0.5 km s$^{\rm -1}$ for HD\,195633. By incorporating these uncertainties, we determined the total absolute uncertainty ($\sigma_{\rm abs}$) for each element in the spectra of the two stars (presented in Table \ref{table:abund}). The $\sigma_{\rm abs}$ for each element is calculated by taking the square root of sum of the square of individual errors associated with each parameter ($T_{\rm eff}$, $\log g$, and $\xi$) for a specific element\footnote{For the uncertainty associated with $\xi$, the abundance variations obtained for a change of +0.5 km s$^{\rm -1}$ were taken into account.}. Following this approach, the $\sigma_{\rm abs}$ values for HD\,8724 range from 0.05 dex (Si\,{\sc i}) to 0.35 dex (Ba\,{\sc ii}). For HD\,195633, $\sigma_{\rm abs}$ values range from 0.06 dex (Si\,{\sc i}) to 0.25 dex (Ba\,{\sc ii}). Details regarding the individual uncertainties for each element can be found in Table \ref{table:abund} (see columns 3 and 7).

No comprehensive analysis of elemental abundances for these two stars has been previously documented in the literature. While several studies have investigated elemental abundances in HD\,8724 and HD\,195633, the work by \cite{mishenina2001} offers the most extensive comparison for a larger number of elements. They reported abundances for seven common elements in both stars using their {\sc ELODIE} spectra. Our current spectroscopic study on HD\,8724 demonstrates good agreement with the findings of \cite{mishenina2001} for the abundance ratios of magnesium ([Mg\,{\sc i}/Fe]), silicon ([Si\,{\sc i}/Fe]), calcium ([Ca\,{\sc i}/Fe]), and barium ([Ba\,{\sc ii}/Fe]) relative to iron (Fe). These abundance ratios differ by $\approx$0.1 dex between the two studies. Similarly, the abundance ratio for yttrium based on two Y\,{\sc ii} lines at 4883.69 \AA\; and 5087.43 \AA\,(see Figure \ref{fig:Y_Zr_Ba}, top panel) shows excellent agreement. For cerium (Ce) and neodymium (Nd), the abundance ratios relative to iron show slightly larger discrepancies.  Specifically, the abundance difference for Ce\,{\sc ii} is about 0.2 dex, while the difference for Nd\,{\sc ii} is around 0.1 dex. The Ce\,{\sc ii} abundance for HD\,8724 and the Solar spectrum is based on a single line at 4628.16 \AA\,(see Figure \ref{fig:Ce_Nd_Sm}, top panel). For Nd, we identified a line at 5293.17 \AA\, in the spectrum of HD\,8724 (see Figure \ref{fig:Ce_Nd_Sm}, middle panel). Finally, a comparison of our element-to-iron abundance ratios for Mg\,{\sc i}, Si\,{\sc i}, Ca\,{\sc i}, Ti\,{\sc i}, Cr\,{\sc i}, Ni\,{\sc i}, Y\,{\sc ii}, and Zr\,{\sc ii} with those reported by \cite{fulbright2000} reveals good agreement within $\approx$0.1 dex.
  
The abundance ratios of Mg\,{\sc i} and Ca\,{\sc i} differ by about 0.1 dex from the values reported for HD\,195633 by \cite{mishenina2001}. The Si abundance ratio is consistent with the findings of \cite{mishenina2001}. However, the abundance ratios for Ba\,{\sc ii} and Y\,{\sc ii} are lower in our study by -0.17 dex and -0.27 dex, respectively (Ba\,{\sc ii} abundances for both stars are based on two lines at 5853.69 \AA\, and 6496.91 \AA, see Figure \ref{fig:Y_Zr_Ba}, bottom panel). \cite{zhang2006} reported abundances for seven common elements in HD\,195633. Compared to their work, our results show generally good agreement with slight differences for individual elements. These differences include -0.13 dex for [Mg\,{\sc i}/Fe], 0.03 dex for [Si\,{\sc i}/Fe], -0.08 dex for [Ca\,{\sc i}/Fe], -0.05 dex for [Ti\,{\sc i}/Fe], -0.04 dex for [Cr\,{\sc i}/Fe], -0.11 dex for [Mn\,{\sc i}/Fe], and -0.01 dex for [Ni\,{\sc i}/Fe], that is, the overall agreement is satisfactory. Our abundance ratios for [Mg\,{\sc i}/Fe], [Si\,{\sc i}/Fe], [Ca\,{\sc i}/Fe], [Ti\,{\sc i}/Fe], and [Ti\,{\sc ii}/Fe] in HD\,195633 differ by $\approx$0.1 dex from those reported by \cite{gratton2003}. However, our titanium abundance ratio based on the neutral (Ti\,{\sc i}) line agrees well with the value reported by \cite{tan2009} (difference of only 0.02 dex). Finally, the zinc (Zn) abundance ratio ([Zn\,{\sc i}/Fe] = 0.25 dex) reported by \cite{roederer2010} is consistent with our findings for HD\,195633.

\section{Age Determination}

\begin{figure*}
    \centering
    \centering\includegraphics[width=0.76\linewidth]{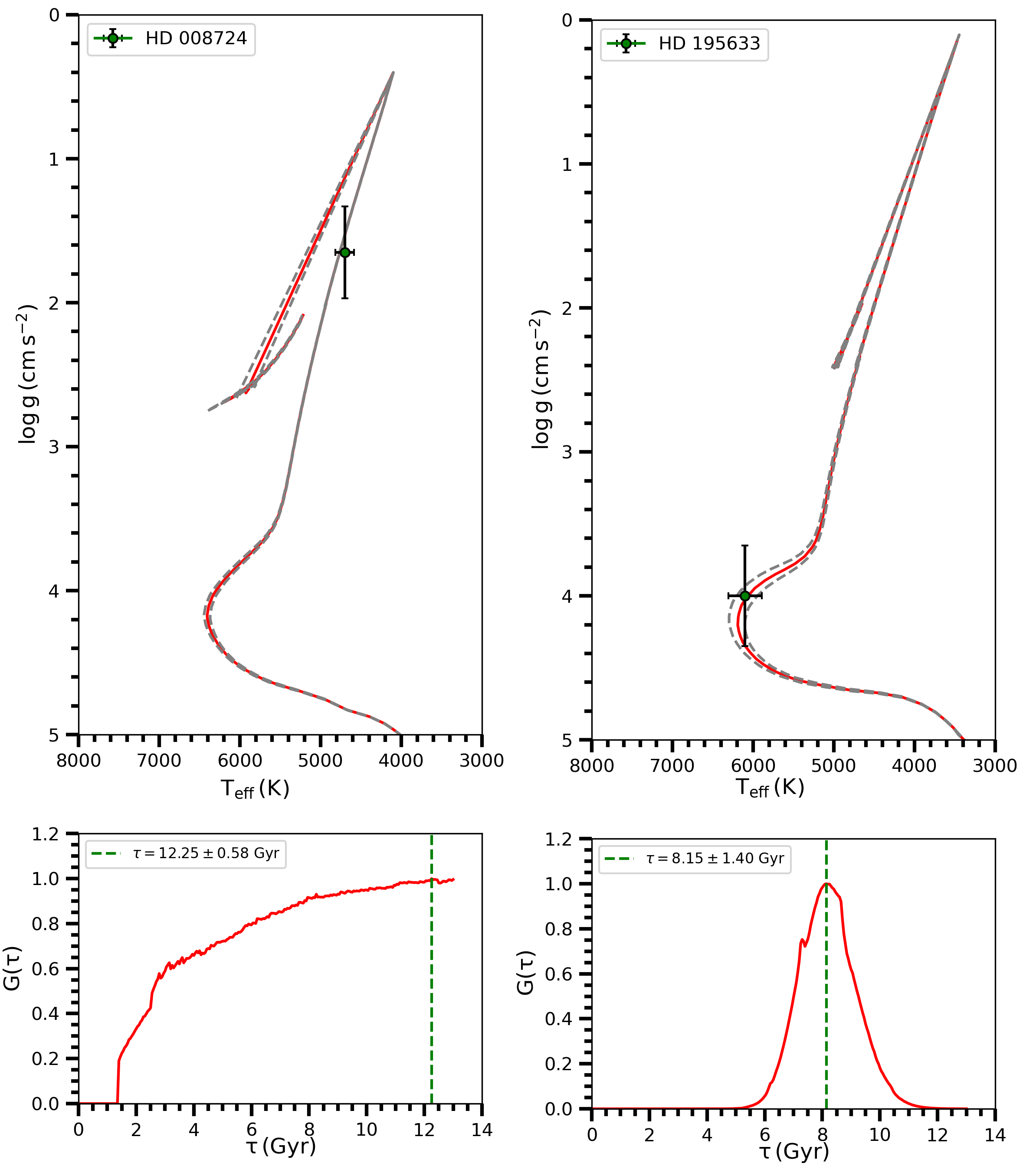}
    \caption{The positions of the stars HD\,8724 and HD\,195633 on the kiel diagrams and the {\sc PARSEC} isochrone contours that best express these positions (upper panels) and the distributions of the probability functions of their ages determined by Bayesian statistics (lower panels).}
    \label{fig:age}
\end{figure*}

The ages of stars HD\,8724 and HD\,195633, for which spectral analyses were performed, were determined by comparing the atmospheric parameters obtained in this study with the stellar evolution model {\sc PARSEC} \citep{Bressan12} using a method based on Bayes statistics developed by \cite{jorgensen2005}. The positions of the stars in the Kiel diagram ($\log g - T_{\rm eff}$) and the best-fitting {\sc PARSEC} \citep{Bressan12} isochrones to Bayes statistics are shown in the upper panels of Figure \ref{fig:age}, whereas the $G$ distributions of the most probable ages calculated using Bayes statistics are shown in the lower panels of the same figure. Details on the calculation of stellar ages can be found in \citet{sahin2020}. By analyzing the positions of the stars in the kiel diagrams, it was observed that HD\,8724 was located in the red-giant branch and HD\,195633 was located above the turning point of the main sequence. According to the method of \cite{jorgensen2005}, the age corresponding to the largest $G$ value is considered the most probable stellar age. According to Bayes statistics, the most probable ages of HD\,8724 and HD\,195633 are determined as $t=12.25\pm0.58$ and $t=8.15\pm1.40$ Gyr, respectively.   

%Figure 6

\section{Kinematic Analyses}
The method developed by \cite{jhonson1987} was used to calculate the space velocity of the two stars analyzed in this study. The right-hand rule was used for analyses. Accordingly, the $U$ space velocity component of the stars was measured in the Galactic center direction, the $V$ space velocity component in the Galactic rotation direction, and the $W$ space velocity component in the north Galactic pole direction. Equatorial  coordinates ($\alpha$, $\delta$), proper-motion components ($\mu_{\alpha} \cos\delta$, $\mu_{\delta}$), trigonometric parallaxes ($\varpi$), and radial velocity ($V_{\rm rad}$) data (see Table \ref{table:6}) were included in the calculations for kinematic analyses. The equatorial coordinates in the 2000 epoch, proper-motion components, and trigonometric parallaxes of the stars  were obtained from the {\it Gaia} DR3 catalog \cite{gaiadr32023}. Radial velocity data were also considered in this study, as determined by the cross-correlation with the solar spectrum. 

\begin{table*}
\setlength{\tabcolsep}{7pt}
\caption{Equatorial coordinates, proper-motion components, trigonometric parallaxes, distances, and radial velocities for two metal-poor stars with HPM. The astrometric and spectroscopic data were taken from \cite{gaiadr32023} catalogue.}
\label{table:6}
\centering
\begin{tabular}{lccccccc}
\hline
Star & $\alpha$& $\delta$ &$\mu_{\rm \alpha} \cos\delta$  &$\mu_{\rm \delta}$ & $\varpi$ & $d$  &$V_{\rm Rad}$\\
%\cline{2-8}
    & (hh:mm:ss) &  (dd:mm:ss) & (mas yr$^{-1}$)  & (mas yr$^{-1}$)   &   (mas)   & (pc) & (km s$^{\rm -1}$)\\	       
\hline
HD\,8724   & 01 26 17.60 & +17 07 35.12 &  52.869$\pm$0.027 & -76.053$\pm$0.017 &  ~~2.310$\pm$0.026 & 433$\pm$5 & -112.98$\pm$0.13\\
HD\,195633 & 20 32 23.99 & +06 31 03.25 &  74.431$\pm$0.024 & 22.554$\pm$0.016  &  10.042$\pm$0.022  & 100$\pm$1 & ~~-45.62$\pm$0.16\\
\hline 
\end{tabular}
\end{table*}

\begin{table*}[h]
\setlength{\tabcolsep}{4pt}
\caption{The calculated space velocity components, total space velocities and obtained Galactic orbital parameters for  HD\,8724 and HD\,195633.}
\label{table:7}
\centering
\begin{tabular}{l|cccc|ccc|c}
\hline
\multirow{2}{*}{Star}  & $U_{\rm LSR}$          &  $V_{\rm LSR}$          & $W_{\rm LSR}$    & $S_{\rm LSR}$   & $Z_{\rm max}$  & $R_{\rm a}$ & $R_{\rm p}$ & $e$ \\
\cline{2-8}
          &  \multicolumn{4}{c|}{(km s$^{\rm -1}$)} & \multicolumn{3}{c|}{(kpc)} &   \\
\hline
HD\,8724    &  ~~~28.31$\pm$0.35  & -204.52$\pm$1.79 & -7.44$\pm$1.04 & 206.60$\pm$2.10  & 0.38$\pm$0.02 & 8.32$\pm$0.01 & 0.28$\pm$0.02 & 0.93$\pm$0.01 \\
HD\,195633  & -46.64$\pm$ 0.11 & ~~~-8.37$\pm$0.12 & ~-1.59$\pm$0.07 & ~~47.41$\pm$0.18 & 0.03$\pm$0.01 & 9.02$\pm$0.01 & 6.47$\pm$0.01 & 0.16$\pm$0.01 \\
\hline 
\end{tabular}
\end{table*}

Space velocity calculations of stars in a Solar neighborhood cannot be performed accurately without considering differential rotation and the local standard of rest (LSR). To eliminate these biases in the stellar space velocity components, a correction (first-order) due to differential rotation was considered. The relations in \cite{mihalas1981} were used for the differential rotation correction of the stars. The differential velocity corrections ($dU, dV$) for stars HD\,8724 and HD\,195633 were calculated as (5.92, -0.52) and (1.99, 0.14) km s$^{\rm -1}$, respectively. Because the differential velocity corrections did not affect the $W$ space velocity component, they were not included in the calculations. Another bias affecting the space velocity components is the spatial space velocity of the Sun calculated with respect to the nearby stars. For this bias, known as the LSR, the value $(U,V,W)_{\odot}=(8.50\pm0.29, 13.38\pm0.43, 6.49\pm0.26)$ km s$^{\rm -1}$ given by \citet{coskunoglu2011} for all Galactic populations is taken. The total space velocities ($S_{\rm LSR}$) of the two stars were calculated by taking the square root of the sum of the squares of their space-velocity components. The space velocities calculated for the two stars are presented in Table \ref{table:7}.

The Galactic orbital parameters of the stars were calculated using {\sc galpy} software developed by \citet{bovy2015}. The code {\sc MWPotential2014}, specially developed for the potential of stellar populations in the Milky Way, was used in the software. In the orbit calculations, the mass and size of the Milky Way components were assumed to be constant over time. 

\begin{figure*}
\centering 
\includegraphics[width=0.49\textwidth]{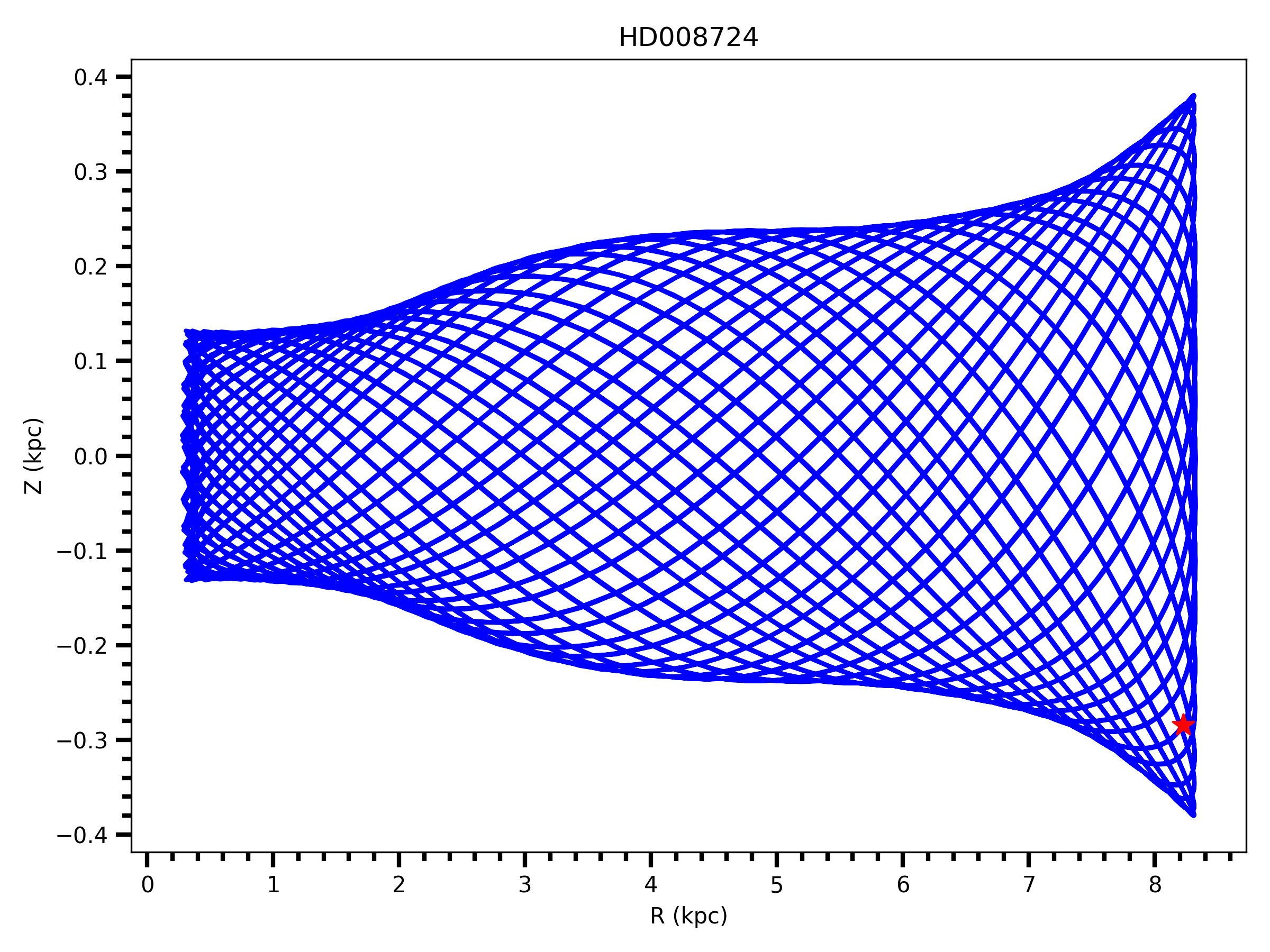}
\includegraphics[width=0.49\textwidth]{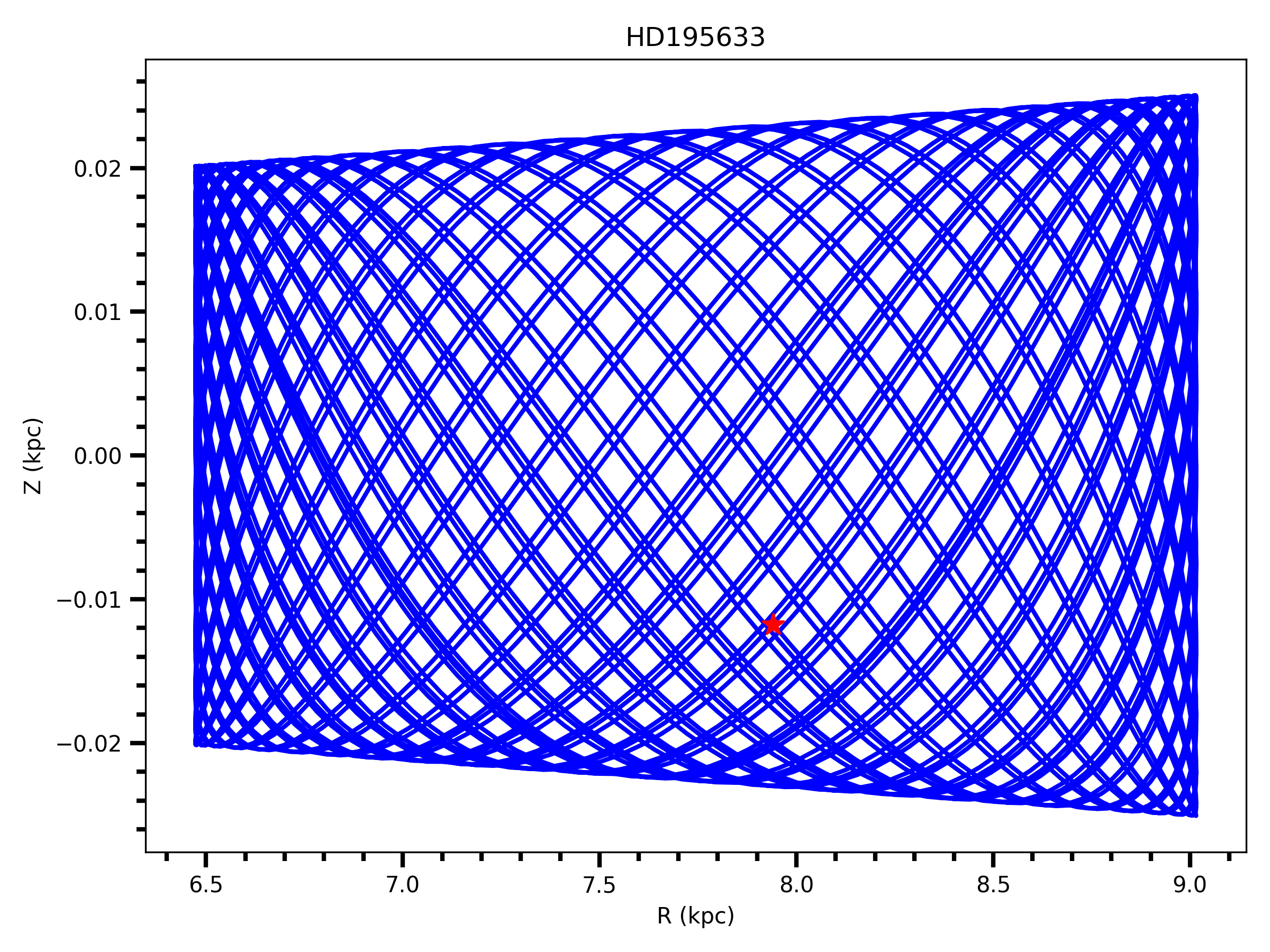}
\caption{The computed meridional Galactic orbits are projected onto the Galactic $Z \times R$ plane for HD\,8724 (left panel), and HD\,195633 (right panel). The red star symbol shows the present day position of the stars in the Galaxy.}
 \label{fig:orb-1}
\end{figure*}

To obtain the precise Galactic orbital parameters of HD\, 8724 and HD\, 195633, the stars were moved around the center of the Galaxy in steps of 650 years at a time interval of 13 Gyr. As a result of the dynamical orbital analyses, the basic parameters of the stars, such as the closest ($R_{\rm p}$) and farthest ($R_{\rm a}$) distances to the Galactic center, the maximum distance they can leave the Galactic plane ($Z_{\rm max}$), and the eccentricities ($e=(R_{\rm a}-R_{\rm p})/(R_{\rm a}+R_{\rm p})$) are calculated and listed in Table \ref{table:7}, and the Galactic orbits of the two stars are shown in Figure \ref{fig:orb-1} in the $Z\times R$ plane. Here, $R$ is the distance between the star and the Galactic center. Considering the Galactic orbits of the stars, it is calculated that HD\, 8724 can reach, $R_{\rm P}=0.28\pm0.02$ kpc, as close as the Galactic center, while HD\,195633 can only reach a distance of $R_{\rm P}=6.47\pm 0.01$ kpc. This is important evidence that HD\, 8724 has a highly flattened orbit ($e=0.93\pm0.01$) as a halo object. In contrast, the orbital eccentricity, $e=0.16\pm0.01$, calculated for HD\,195633 indicates that the object belongs to the Galactic disc \citep{plevne2015, onal2018, tasdemir2023}.

%Figure 7

\section{Dynamical Orbital Analyses, Galactic Origins of Stars and Concluding Remarks}

\citet{sahin2020} showed that the presence of high proper-motion metal-poor stars in the vicinity of the Sun can be explained by a scenario in which these stars break away from the GCs. To investigate the Galactic origin of the two stars analyzed in this study, the Galactic orbital parameters and orbits of the GCs in the Milky Way were constructed by considering equatorial coordinates, distance, proper motion, and radial velocity data collected from the literature \citep{Baumgardt2019,vasiliev2021}. Based on the data collected for the GCs, kinematic and dynamic calculations were performed for 170 known GCs on the Milky Way.

As mentioned in Section 4, the {\sc MWPotential2014} potentials developed by \citet{bovy2015} for the Milky Way in the {\sc galpy} Python library were used to calculate the orbital parameters of GCs. In the orbital analyses of GCs, orbital motions consisting of 20 million points from the present day to 13 Gyr in the past were considered. The distance between the two points in this orbital motion was 650 yr, and this value was determined after several trials considering the computation time and processing capacity. The 20 million-point orbital motion is sufficiently precise to capture the stellar-GC encounter and can be computed in a relatively short time interval using multi-core processors. 

The equatorial coordinates, proper-motion components, trigonometric parallaxes, and radial velocities required for the orbital calculations of HD\,8724 and HD\,195633 stars were obtained from the {\it Gaia} DR3 database \citep{gaiadr32023}. The method used for the two stars here investigated is also applied to stars in GCs. Consequently, the position and velocity values of the orbits of the stars were calculated from the present day to 13 Gyr in the past, as stated in the calculation of the orbits of the clusters. Although the synchronous orbits were calculated 13 Gyr backward from the present, the probabilities of encounter for each star were calculated for the orbital points after birth. Points before birth were not included in the probability calculation.

By calculating the synchronous orbits of HD\,8724 and HD\,195633, the simultaneous presence of these stars at a distance of five tidal radii from the centers of approximately 170 GCs \citep{vasiliev2021} was examined. Accordingly, the distance between the orbital positions of the objects was calculated for each time step. For the time step corresponding to the time step in which the results obtained were less than five tidal radii, an encounter was assumed to exist, and the GC-star encounter  position ($\Delta \theta$) and relative velocity ($\Delta \nu$) were calculated using the following equations: 

\begin{equation}
    \Delta \theta = \sqrt{(X_{\rm s}-X_{\rm GC})^2+(Y_{\rm s}-Y_{\rm GC})^2+(Z_{\rm s}-Z_{\rm GC})^2},
\end{equation}

\begin{equation}
    \Delta \nu = \sqrt{(U_{\rm s}-U_{\rm GC})^2+(V_{\rm s}-V_{\rm GC})^2+(W_{\rm s}-W_{\rm GC})^2}.
\end{equation}
Here $\Delta \theta$ is the distance difference between the star (s) and the globular cluster (GC), defined in the Cartesian system, at a given time $t$. Similarly, $\Delta \nu$ is the velocity difference between the space velocity components of the star and the GCs. $X$, $Y$, and $Z$ are the positions of the objects in the Cartesian system and $U$, $V$, and $W$ are the space velocity components of the objects. 

When a GC-star encounter occurred at more than one time step, the probability value for that time section was determined from the sum of the probabilities of the successive encounters. Additionally, in the case of encounters at more than one time step, the highest probability value of the time step was selected for the cluster and used for comparison with other clusters. The probabilities of the position ($P(\theta)$) and velocity ($P(\nu)$) differences calculated over a wide time interval, under the assumption that they have a Gaussian distribution, were calculated using Equations \ref{Eq: Eq1} and \ref{Eq: Eq2}, respectively.

% Table 8
\begin{table*}[htbp]
  \centering
  \caption{Data for the five most likely GCs that support the GC escape scenario for HD\,8724 and HD\,195633.}
    \begin{tabular}{lccccccccc}
    \hline
    Cluster & $l$ & $b$ & $P({\rm \theta})$  & $P({\rm \nu})$  & $P({\rm origin | \theta, \nu})$ &  [Fe/H] & $\tau$ & [Mg/Fe] & References\\
       & ($^{o}$) & ($^{o}$) & (\%)  & (\%)  & (\%) &  (dex) & (Gyr) & (dex) &  \\ 
    \hline
    NGC\,5139 & 309.10 & 14.97  & 89 & 66    & 59    &  -1.65 & 13     & 0.43$\pm$0.22$^{\rm*}$ & 01, 01, 02 \\
    Terzan\,5 & ~~~3.84& ~~1.69 & 88 & 65    & 57    &  -0.25$\pm$0.20 & 12$\pm$1      & 0.33$\pm$0.10 & 03, 04, 06 \\
    NGC\,6441 & 353.53 & ~~5.01 & 81 & 64    & 52    &  -0.44$\pm$0.07 & 11.2$\pm$2.4  & 0.11$\pm$0.06 & 06, 07, 08 \\
    NGC\,6316 & 357.18 & ~~5.76 & 91 & 55    & 50    &  -0.86$\pm$0.02 & 14.83$\pm$0.93& 0.11$\pm$0.06 & 09, 09, 08 \\
    Terzan\,9 & ~~~3.60& ~-1.99 & 75 & 66    & 49    &  -1.10$\pm$0.15 & 13            & 0.27$\pm$0.03 & 10, 10, 10 \\
\hline
    HD\,8724  & 134.74 & -44.94 & -- & --    & --    &  -1.59$\pm$0.04 & 12.25$\pm$0.58& 0.35$\pm$0.05 & 11, 11, 11 \\
\hline \hline
    NGC\,5139 & 309.10 &  14.97 & 96 & 71    & 69   &  -1.65           & 13             & 0.43$\pm$0.03 & 01, 01, 02 \\
    NGC\,7078 & ~~64.01& -27.31 & 63 & 91    & 58   &  -2.22$\pm$0.14  & 13.6           & 0.41$\pm$0.03 & 12, 12, 08 \\
    NGC\,6656 & ~~~9.89&~~-7.55 & 70 & 81    & 57   &  -1.52$\pm$0.09  & 12.7           & 0.50$\pm$0.01 & 12, 12, 08 \\
    NGC\,2808 & 282.19 & -11.25 & 60 & 78    & 46   &  -0.92$\pm$0.07  & 11.2           & 0.22$\pm$0.04 & 12, 12, 08 \\
    NGC\,6356 & ~~~6.72& ~10.21 & 42 & 85    & 36   &  -0.35$\pm$0.14  & 11.35$\pm$0.41 & 0.12$\pm$0.04 & 13, 13, 08 \\
    \hline
    HD 195633 &~~62.37& 48.27  & --    & --    & --   &  -0.52$\pm$0.05 & 8.15$\pm$1.40  & 0.07$\pm$0.05 & 11, 11, 11 \\
    \hline
    \end{tabular}%
  \label{tab:final}%

\begin{minipage}{17cm}
\vskip 4pt
[01] \citet{villanova2014}, [02] \citet{magurno2019}, [03] \citet{massari2014}, [04]\citet{ferraro2016}, [05] \citet{origlia2011}, [06] \citet{carretta2009b}, [07] \citet{marin-franch2009}, [08] \citet{dias2016}, [09] \citet{cezario2013}, [10] \citet{ernandes2019}, [11] This study, [12] \citep{kovalev2019}, [13] \citep{koleva2008}, (*) The reported dispersion in [Mg/Fe] ratio by \citet{magurno2019}.
\end{minipage}
  \end{table*}%

\begin{equation}
    P({\rm \theta}) = \frac{1}{\sqrt{2\pi R_{\rm tidal}}}\times \exp\left(-\frac{(\Delta \theta)^2}{2\times R_{\rm tidal}}\right).
    \label{Eq: Eq1}
\end{equation}

\begin{equation}
    P({\rm \nu})= \frac{1}{\sqrt{2\pi V_{\rm escape}}}\times \exp\left(-\frac{(\Delta \nu)^2}{2\times V_{\rm escape}}\right),
    \label{Eq: Eq2}
\end{equation}
Here, $R_{\rm tidal}$ is five times the cluster tidal radius and $V_{\rm escape}$ is the escape velocity of the GC, and their values are taken from 4th version of the Milky Way GC Database\footnote{\href{https://people.smp.uq.edu.au/HolgerBaumgardt/globular/}{https://people.smp.uq.edu.au/HolgerBaumgardt/globular/}}. In the combined evaluation of the spatial and velocity probabilities calculated for each GC, the multiplication of the two probability values ($P({\rm origin | \theta, \nu})$) was adopted, and the following equation was used:   
\begin{equation}
    P({\rm origin | \theta, \nu})= P({\rm \theta}) \times P({\rm \nu}).
\end{equation}

%Figure 8
\begin{figure}
    \includegraphics[width=0.52\textwidth]{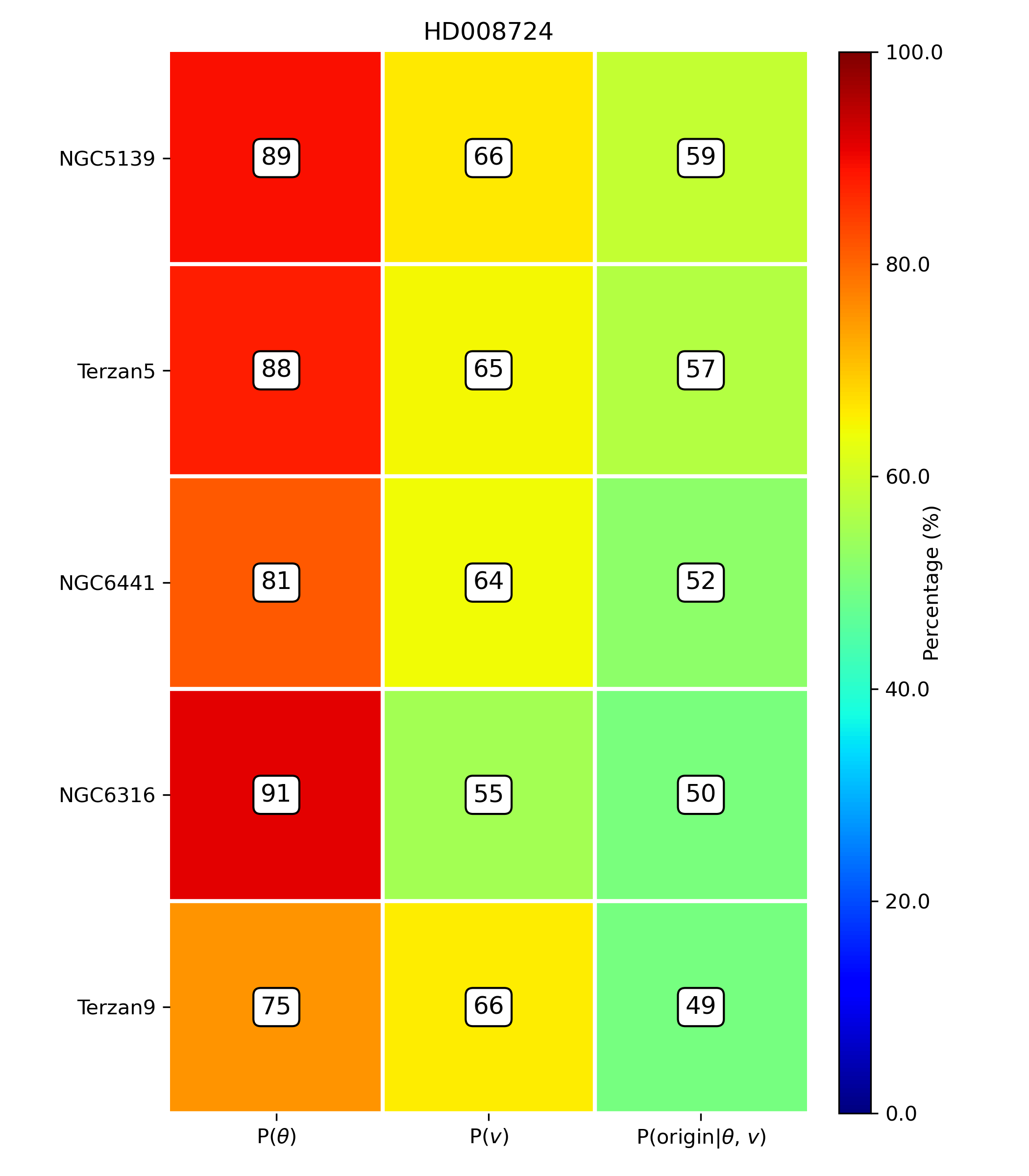}
    \includegraphics[width=0.52\textwidth]{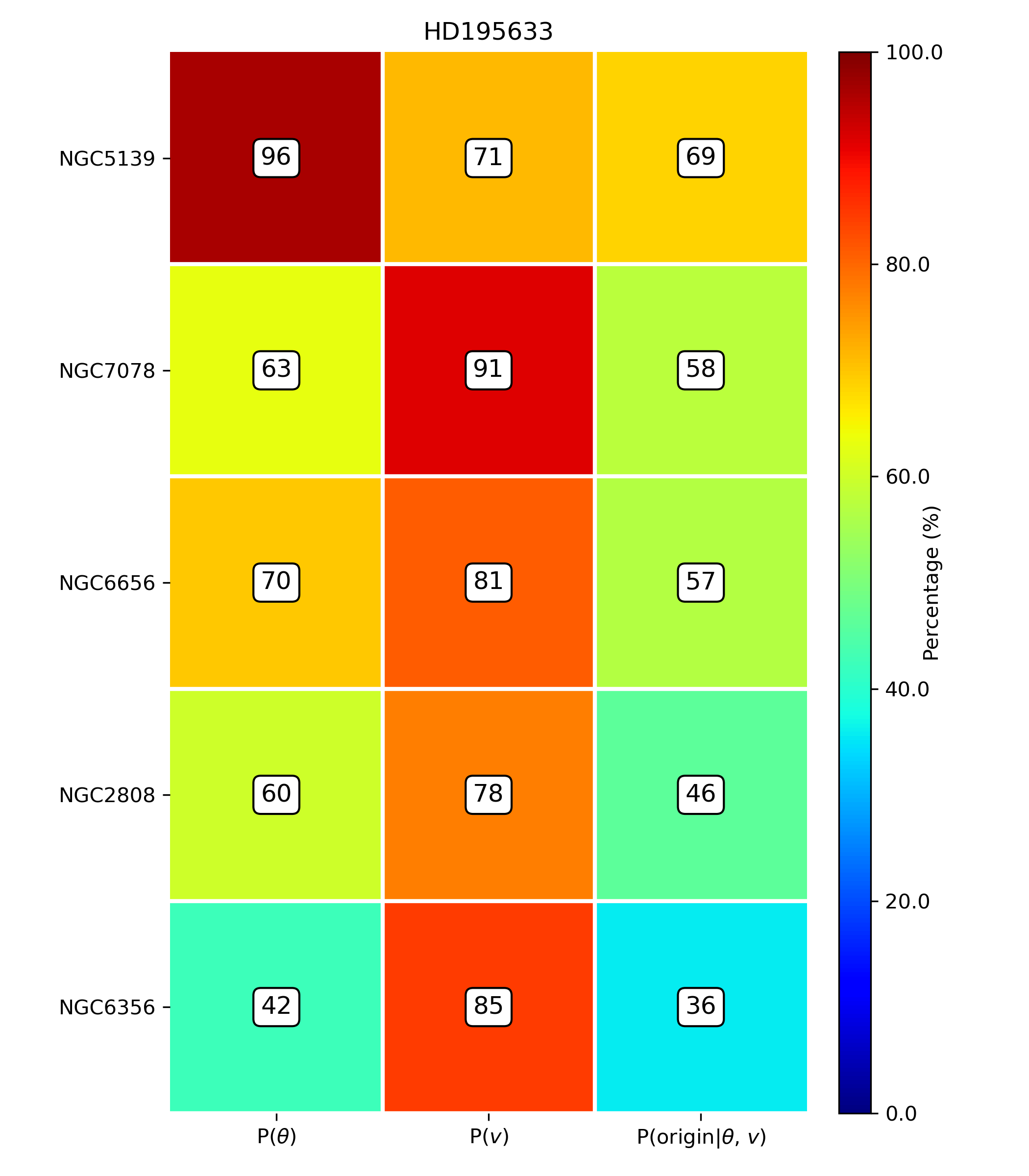}
    \caption{Probability matrices for spatial and velocity encounters between stars and GCs. Each panel contains three probability values for the five GCs that a star is most likely to encounter: position ($P({\rm \theta})$), velocity ($P({\rm \nu})$) and the product of these two probabilities ($P({\rm origin| \theta, \nu)}$). The colour of each cell is displayed according to the calculated probability values, depending on the colour scale given on the right of the panels.}
    \label{fig:origin}
\end{figure}

%Figure 9
\begin{figure}
    \centering
                \includegraphics[width=0.48\textwidth]{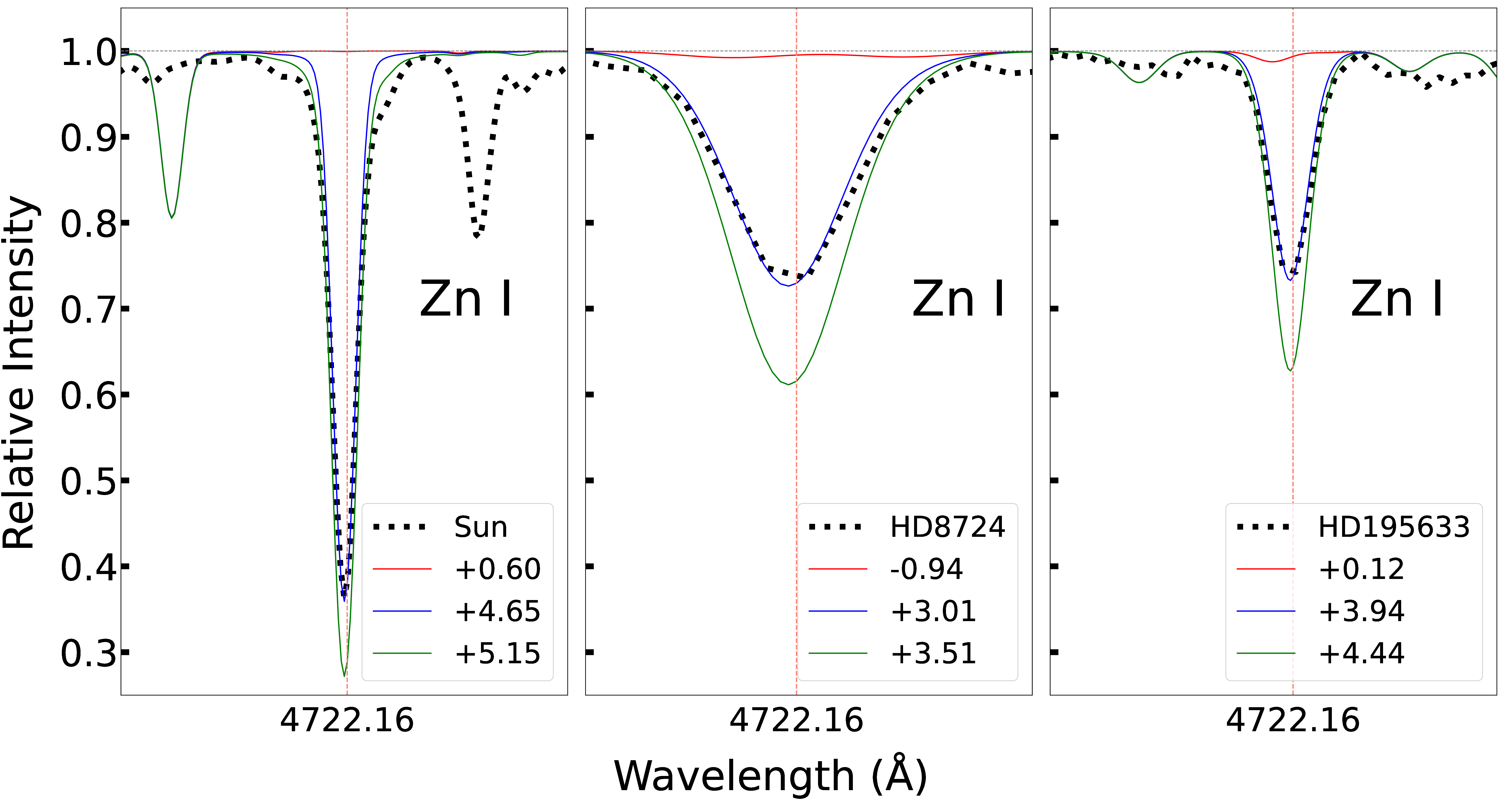}
                    \includegraphics[width=0.48\textwidth]{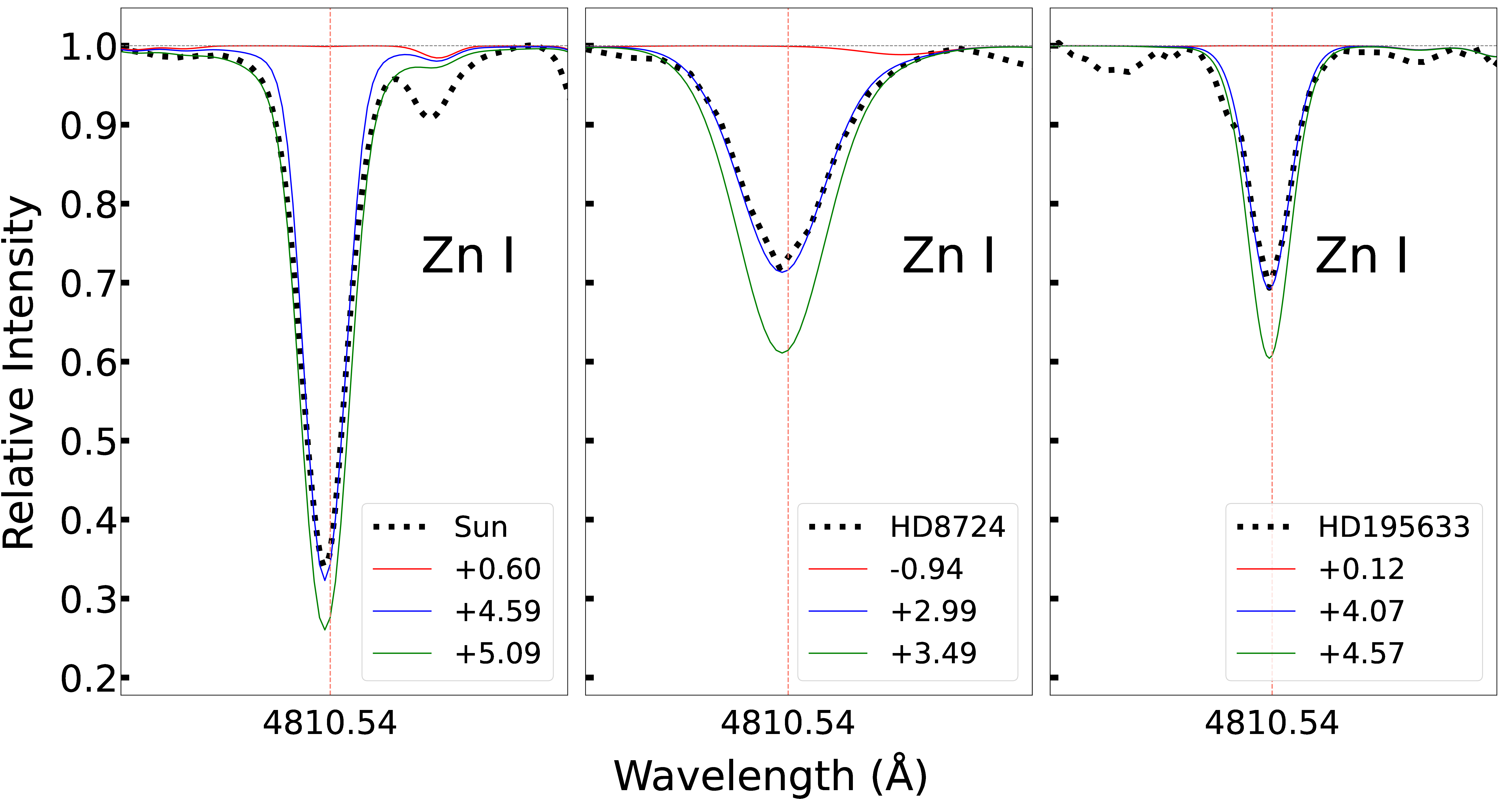}
    \caption{The observed (filled circles) and computed (full blue line) line profiles for the neutral zinc lines at 4722.16 \AA\, and 4810.54 \AA\,used in the analysis of HD\,8724, HD\,195633 and the Sun. The computed profiles illustrate the synthetic spectra for the three varying logarithmic abundances. The red dotted lines are the spectra computed with no contribution from Zn\,{\sc i}.}
    \label{fig:Zn1}
\end{figure}

The probabilities calculated from the position and velocity comparison of HD\,8724 and HD\,195633 with 170 GCs in the Milky Way \citep{vasiliev2021} and the probability values, including their combination, are shown in Figure \ref{fig:origin} with probability matrices including the five most probable GCs. Table \ref{tab:final} provides the encounter probabilities that serve as an alternative diagnostic tool for evaluating the dynamic origins of program stars. The probabilities were calculated from the kinematic analyses of stars and GCs. Taking into account the scenario of stars escaping from GCs, we compared the position and velocity of the two stars here analyzed with GCs, and we found that HD\,8724 and HD\,195633 had encounter probabilities of $49\%\leq P({\rm origin})\leq 59\%$ and $36\% \leq P({\rm origin})\leq69\%$, respectively. However, these encounter probabilities alone are insufficient for determining the kinematic origins of the stars. In addition to the time of encounter, the age and metal abundances of the stars were expected to be consistent with those of candidate GCs. To identify candidate GCs for the escape scenario, the iron ([Fe/H]) and magnesium ([Mg/Fe]) abundances and ages ($\tau$) of the selected GCs were compiled from literature (Table \ref{tab:final}). 

%Figure 10
\begin{figure}
    \centering
    \includegraphics[width=0.48\textwidth]{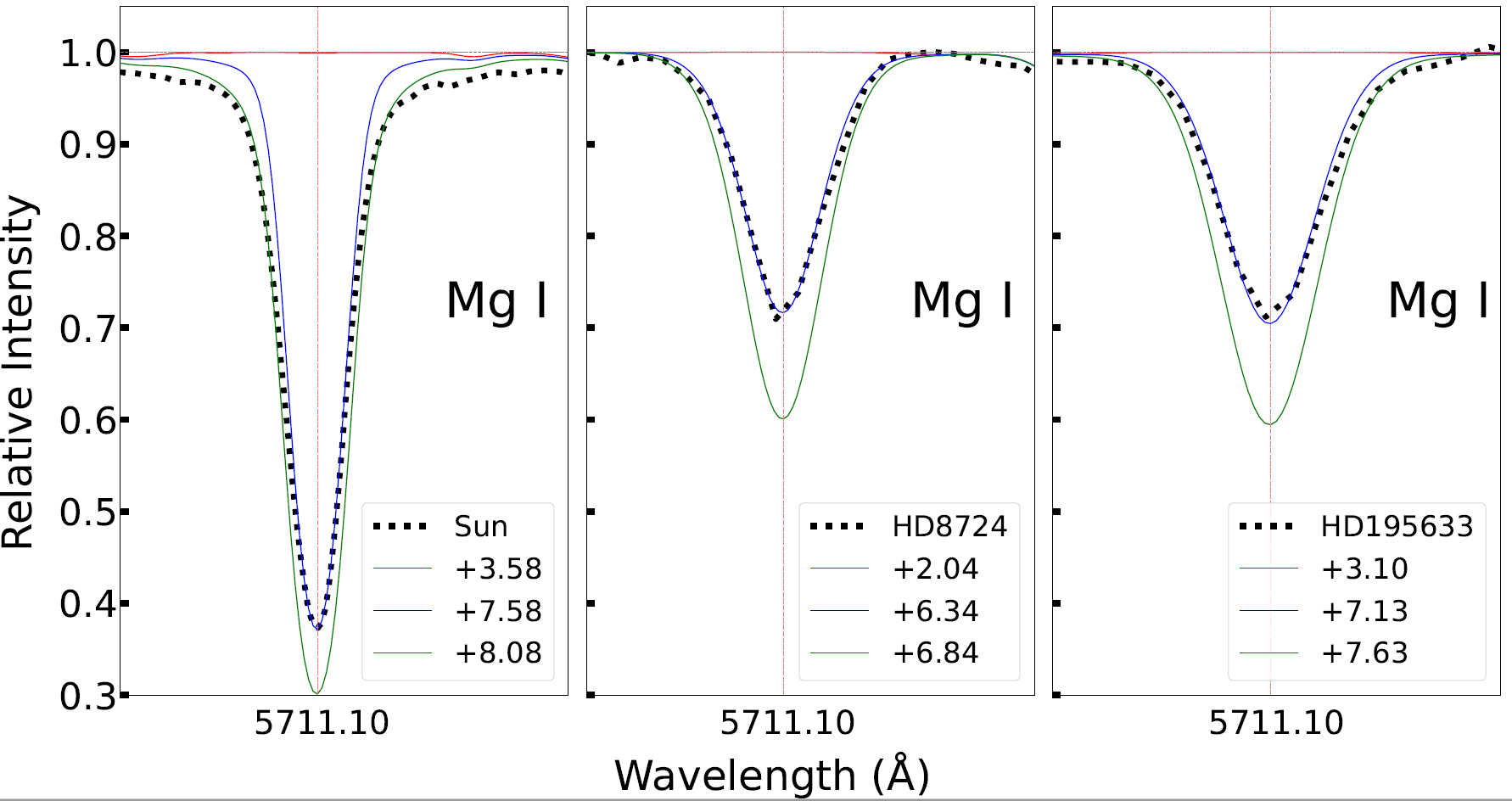}
    \caption{The observed (filled circles) and computed (full blue line) line profiles for the neutral magnesium line at 5711.10 \AA\, used in the analysis of HD\,8724, HD\,195633 and the Sun. The computed profiles illustrate the synthetic spectra for the three varying logarithmic abundances. The red dotted lines are the spectra computed with no contribution from Mg\,{\sc i.}}
    \label{fig:Mg}
\end{figure}

\begin{figure*}
    \centering
    \includegraphics[width=0.95\textwidth]{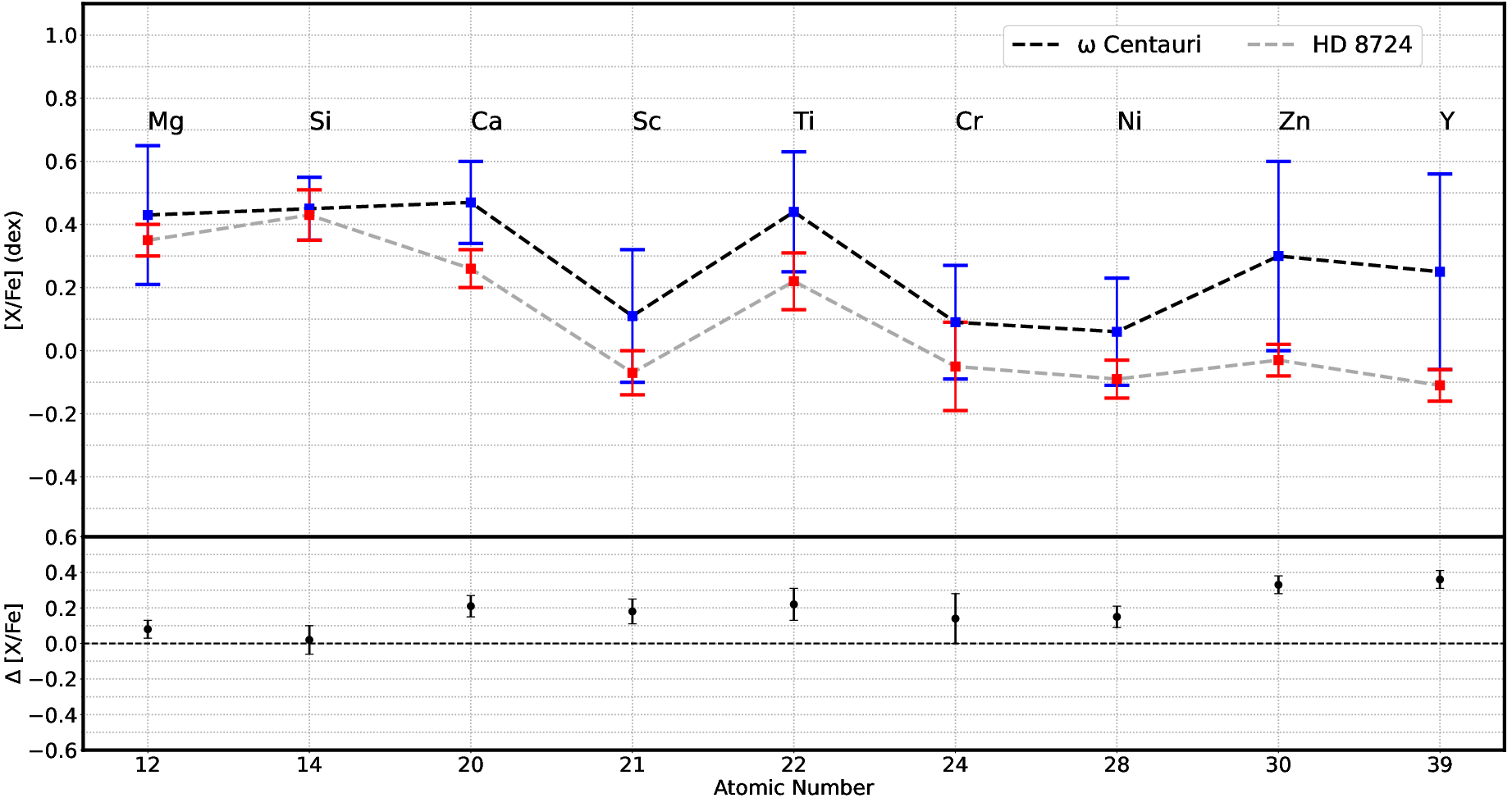}
    \caption{Abundances of NGC\,5139 ($\omega$ Cen) along with the abundances of HD\,8724.}
    \label{fig:ngc5139}
\end{figure*}

NGC\,5139 ($\omega$ Cen) seems to be the most likely candidate among the five possible GCs identified for membership of HD\,8724. Kinematic analyses in this study showed that HD\,8724 may belong to NGC\,5139 with a probability of $P({\rm origin})=59\%$. The metal abundance and age of the GC were in excellent agreement with the metal abundance and age of the star reported in this study \citep{villanova2014, magurno2019}. Figure \ref{fig:ngc5139} presents abundances of HD\,8724 and GC candidate NGC\,5139. The agreement in abundances is satisfactory. \citet{magurno2019} do not report the silicon abundance for NGC\,5139. The [Si/Fe] abundance was provided from \citet{garay2024}. \citet{garay2024} reported the Fe, Al, Mg and Si abundances of 439 red giant branch (RGB) stars in the $\omega$ Cen and identified four distinct Fe populations for the cluster. The metallicity value reported in the study for the population group closest to the metallicity of HD\,8724 was [Fe/H] = -1.55$\pm$0.10 dex (153 stars). In their study, using the dispersion of this metallicity value (0.10 dex) as a criterion, we identified 102 RGB stars among the 439 RGB stars analyzed by \citet{garay2024} that fall in the metallicity range -1.45$<$[Fe/H](dex) $<$-1.65, and calculated the 1/$\sigma^{\rm 2}$ weighted averages for the [Mg/Fe] and [Si/Fe] abundances for the GC over the sample group. The calculated metallicity value for these 102 RGB stars is [Fe/H]=-1.55$\pm$0.06 dex. Accordingly, the ratios [Mg/Fe] = 0.25$\pm$0.23 dex and [Si/Fe]=0.45$\pm$0.10 dex for these 102 RGB stars were in excellent agreement with the ratios reported in this study for HD\,8724 in Table \ref{table:abund}.  

%Figure 11

The main sequence of $\omega$ Cen was divided into two distinct parallel sequences, each exhibiting different metallicities. The dissimilar metallicity values suggest diverse helium content \citep{bedin2004, piotto2005}. In other words, the stars in this massive GC can be categorized into metallicity groups that span a large range \citep[$-2.2 \leq {\rm [Fe/H] (dex)} \leq -0.6$;][]{johnson2010, marino2011, villanova2014}. These sequences had moderately varied ages. In some instances, differences of 2-4 Gyrs have been reported (\cite{Ferraro2004}; \cite{Freyhammer2005}. \citet{Lee2005} estimated a difference of $\approx$1.5 Gyr using a similar hypothesis regarding the helium enhancement. \cite{Stanford2006} concluded that the most likely age difference in $\omega$ Cen was 2-4 Gyrs. A comprehensive list of the age spreads reported in the literature for $\omega$ Cen can be found in Table 5 of \citet{Stanford2006}. Regarding metallicity, \citet{sollima2005} reported four stellar populations at [Fe/H] = -1.7 dex, -1.3 dex, -1.0 dex, and -0.6 dex. \citet{villanova2007} found three stellar populations at [Fe/H] = -1.68 dex, -1.37 dex, and -1.14 dex. \cite{Calamida2009} found six peaks in the iron distribution at [Fe/H] = -1.73 dex, -1.29 dex, -1.05 dex, -0.80 dex, -0.42 dex, and -0.07 dex. \citet{johnson2010} identified four groups with [Fe/H] = -1.75 dex, -1.50 dex, -1.10 dex, and -0.75 dex. \cite{marino2011} reported clear peaks at [Fe/H] = -1.76 dex, -1.60 dex, -1.00 dex, and -0.76 dex. \citet{villanova2014} identified six sub populations at [Fe/H] = -1.83 dex (pop$_{\rm 1}$), -1.65 dex (pop$_{\rm 2}$), -1.34 dex (pop$_{\rm 3}$), -1.05 dex (pop$_{\rm 4}$), -0.78 dex (pop$_{\rm 5}$), and -0.42 dex (pop$_{\rm 6}$). With a metallicity of [Fe/H] = -1.59 $\pm$ 0.04 dex, HD\,8724 may belong to the pop$_{\rm 2}$ sub population of $\omega$ Cen defined by \citet{villanova2014}. A very recent study by \citet{nitschai2024} involving 11\,050 stars in the  GC NGC\,5139 reported a median metallicity value of [M/H] = -1.614$\pm$0.003 dex for the cluster. This value is consistent within 0.02 dex of the metallicity value reported for HD\,8724 in this study. The same study used a Gaussian mixture model to determine the metallicity distribution and reported a stellar fraction of 13.8$\%$ for cluster stars with a metallicity of [M/H] = -1.553$\pm$0.036 dex. This stellar fraction ratio indicates that stars with similar metallicities to HD\,8724 are relatively abundant within the cluster. The $\alpha$-elements for $\omega$ Cen exhibited a slight enhancement, as expected for old stars, with a mean abundance of [$\alpha$/Fe] = 0.41$\pm$0.02 dex \citep{magurno2019}. The abundances\footnote{The errors on abundances are dispersion values reported by \citet{magurno2019} in their Table 8.} for iron peak elements Sc, Cr, and Ni reported by \citet{magurno2019} displayed nearly solar abundances ([Sc/Fe] = 0.11$\pm$0.21, [Cr/Fe]= 0.09$\pm$0.18, [Ni/Fe]= 0.06$\pm$0.17), with the exception of zinc ([Zn\,{\sc i}/Fe]= 0.30$\pm$0.11 dex), which appeared to be slightly enhanced. In contrast, the abundance of the s-process element Y exhibited peculiar characteristics, suggesting the presence of two distinct populations within the $\omega$ Cen. The more metal-rich tail ([Fe/H] $\geq$ -1.5 dex) consisted mainly of stars with strong s-process enrichment, as evidenced by the mean abundance of [Y/Fe]$\geq 0.4$ dex, whereas the more metal-poor stars had nearly solar abundances. The mean cluster abundance of yttrium ([Y/Fe]) is 0.25$\pm$0.31 dex \citep{magurno2019}. The Sc, Cr, and Ni abundances reported by \citep{magurno2019} are in good agreement with those reported for HD\,8724 in this study, i.e., [Sc\,{\sc ii}/Fe] = -0.07$\pm$0.07 dex, [Cr\,{\sc ii}/Fe] = 0.05$\pm$0.11 dex, [Ni\,{\sc i}/Fe] = -0.09$\pm$0.06 dex, respectively. The nickel abundance was provided by 20 neutral Ni lines. Zn abundance was an exception. Figure \ref{fig:Zn1} shows the synthetic spectra for the Zn\,{\sc i} lines at 4722.16 \AA\,and 4810.54 \AA. The abundance for 4722.16 \AA\, Zn\,{\sc i} line gives almost the same logarithmic abundance with Zn\,{\sc i} line at 4810.54 \AA, i.e., the difference is -0.02 dex. In contrast, in $\omega$ Cen, Zn\,{\sc i} lines were only observed in a handful of stars, especially in the metal-rich tail of the sample of \citet{magurno2019}, and the Zn abundance from those stars may not be representative for the Zn abundance of the cluster. The yttrium abundance ([Y\,{\sc ii}/Fe] = -0.08$\pm$0.14 dex) for HD\,8724 can be accepted as nearly solar. Also, Mg abundance is a critical input parameter used to evaluate cluster escape scenarios. Therefore, it is extremely important to accurately calculate magnesium abundance. To this end, the elemental abundance of the 5711.10 \AA\, Mg\,{\sc i} line was obtained using the spectrum synthesis technique (Figure \ref{fig:Mg}). 

Similar kinematic analyses for HD\,195633 showed that, with a 69$\%$ probability of $P({\rm origin})$, the star could be a member of GC NGC\,5139. However, the metallicity of -1.65 dex, the [Mg/Fe] ratio of 0.43 dex and the age of 13 Gyr reported for the cluster indicate that a membership assessment for the star based on kinematic analyses alone cannot be correct (\citealp{villanova2014, magurno2019}). Considering the kinematic analyses, metal abundances, and ages of  NGC\,7078 ($P({\rm origin})=58\%$), NGC\,6656 ($P({\rm origin})=57\%$), and NGC\,2808 ($P({\rm origin})=46\%$) GCs from which the star is likely to have escaped, it seems unlikely that HD\,195633 was separated from these GCs because of the large differences between the metal abundances and ages of the three clusters. In contrast, the values reported for the NGC\,6356 cluster, for which a membership probability of 36$\%$ was calculated, are in agreement with the values obtained for the star (see Table \ref{tab:final}). 

The [Fe/H] and [Mg/Fe] abundances of NGC\,6356 were within the error limits of those reported in this study for HD\,195633 \citep{koleva2008, dias2016}. However, age was an exception, unless NGC\,6356 was assumed to contain sub populations with different metallicities, such as NGC\,5139. NGC\,6356 is known as a bulge GC \citep{minniti1995} and bulge GCs, such as NGC\,6388, a high-mass and high-metallicity GC, are known to possess multiple stellar populations with variations in light elements, including O, Na, Mg, and Al \citep{Carretta2023}. Similarly, \citet{Ferraro2009} reported that Terzan 5, another bulge GC, exhibited two distinct stellar populations with varying iron contents and ages, which may indicate a complex formation history. These findings align with the broader understanding that GCs often host stars of different generations, as evidenced by the presence of multiple populations (MPs) in 14 GCs towards the southern Galactic bulge \citep{Kader2022}. Moreover, the analysis of seven GCs in the Galactic bulge has demonstrated that they can also be consistently aged and host first- and second-generation stars with minimal age differences \citep{Oliveira2020}. 

For NGC\,6356, the iron abundances ([Fe/H]) reported by \citet{dias2016} for 13 cluster member stars range from -1.11 dex to 0.02 dex. The mean [Mg/Fe] ratio for this sample of stars, $0.16\pm$0.09 dex, is within the limits of error of the magnesium abundance ([Mg/Fe] = $0.07\pm$0.05 dex) reported in this study for HD\,195633. In summary, similar to other bulge GCs, NGC\,6356 might contain multiple stellar populations. However, without specific data on NGC\,6356, it is impossible to conclude definitively that MPs of different ages are present in this particular cluster. Further spectroscopic and photometric studies are required to confirm these features, specifically for NGC\,6356.
 
\section{Acknowledgements}
The authors express their sincere gratitude to the reviewer Dr. Elisabetta Caffau for providing invaluable feedback and suggestions that have significantly enhanced the readability and overall quality of the paper. This study was supported by \fundingAgency{Scientific and Technological Research Council of Turkey (TUBITAK)} under the Grant Number \fundingNumber{121F265}. The authors thank to TUBITAK for their supports. This work has made use of data from the European Space Agency (ESA) mission \emph{Gaia}\footnote{https://www.cosmos.esa.int/gaia}, processed by the \emph{Gaia} Data Processing and Analysis Consortium (DPAC)\footnote{https://www.cosmos.esa.int/web/gaia/dpac/consortium}. Funding for DPAC has been provided by national institutions, in particular, the institutions participating in the \emph{Gaia} Multilateral Agreement. This research made use of NASA’s Astrophysics Data System and the SIMBAD database, operated at CDS, Strasbourg, France. The non-public data underlying this article will be made available upon reasonable request from the authors.

\subsection*{Author contributions}

\textbf{Conception/Design of study}: TS, MM, SB;\\ 
\textbf{Data Acquisition}: MM, TS, FG, OP; \\
\textbf{Data Analysis/Interpretation}: TS, MM, FG, SB, OP; \\
\textbf{Drafting Manuscript}: TS, MM, FG; \\
\textbf{Critical Revision of Manuscript}: TS, SB; \\
\textbf{Final Approval and Accountability}: TS, MM, SB, OP.

\subsection*{Financial disclosure}

None reported.

\subsection*{Conflict of interest}

The authors declare no potential conflict of interests.

%\nocite{*}% Show all bib entries - both cited and uncited; comment this line to view only cited bib entries;
%\bibliography{Wiley-ASNA}%

\appendix
%Figure 3

\section{Comparison of model atmosphere parameters for HD\,8724 and HD\,195633}
%Figure A1
\begin{figure*}
    \centering
    \includegraphics[width=0.97\textwidth]{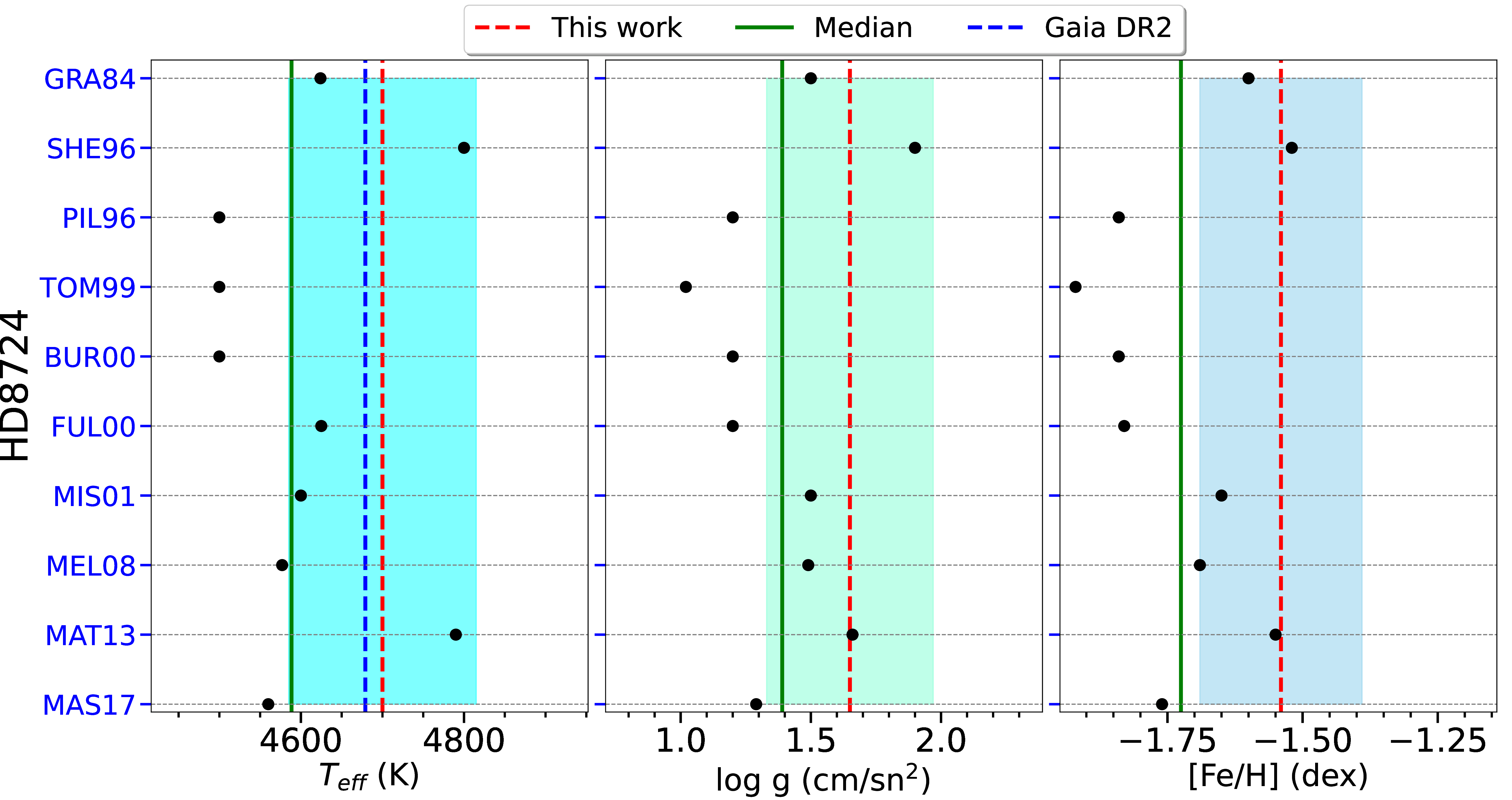}
        \includegraphics[width=0.97\textwidth]{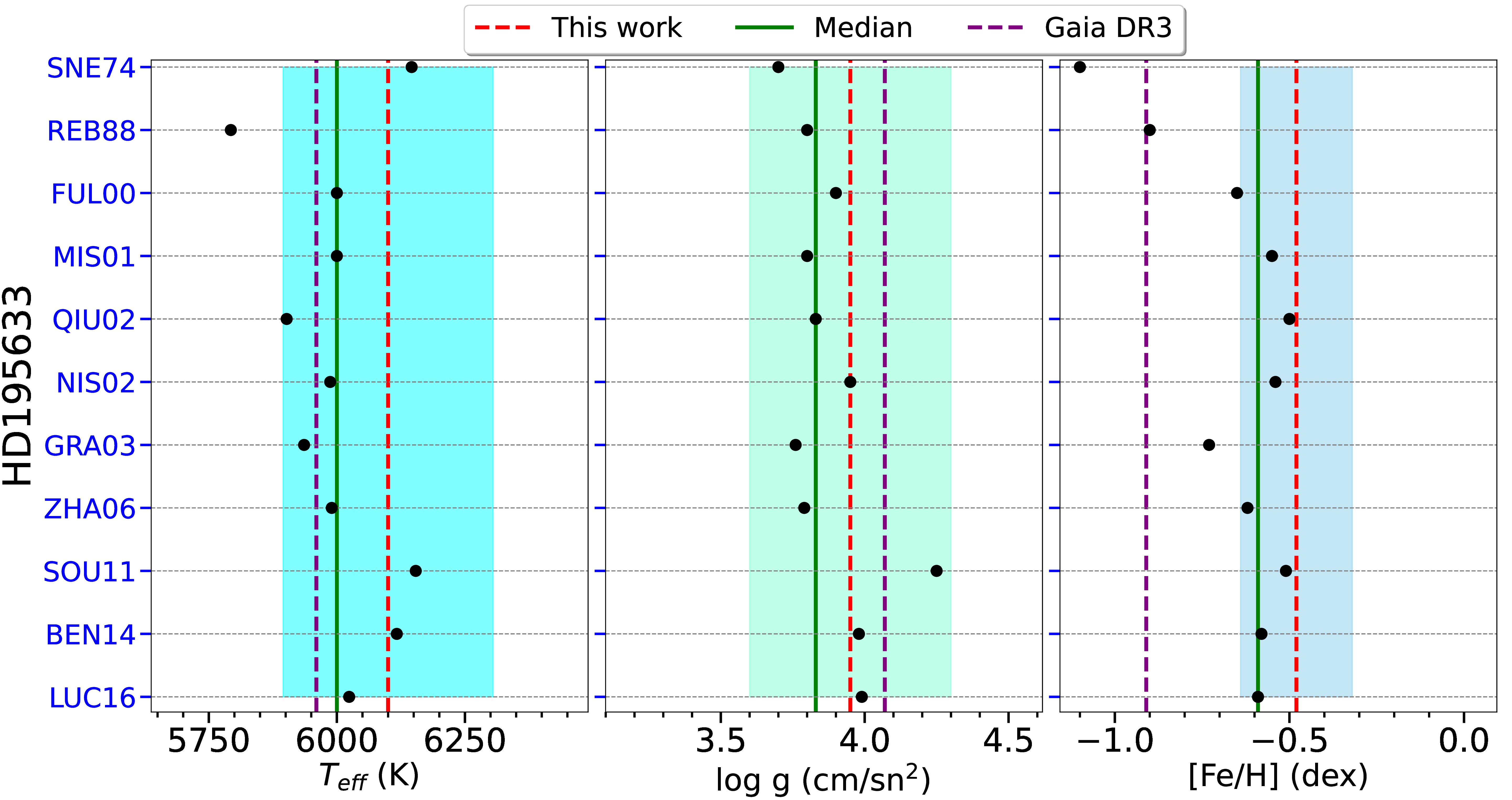}
    \caption{Comparison of model atmosphere parameters for HD\,8724 (upper panel) and HD\,195633 (lower panel). References for the stellar model parameters reported in the PASTEL catalog \citep{soubiran2016} for stars are shown on the y-axis. In all panels, the stellar model parameter values obtained in this study are indicated by the dashed red line. The blue dashed lines indicate the {\it Gaia} DR2 values \citep{gaiadr22018} (upper panel) and {\it Gaia} DR3 values \cite{gaiadr32023} (lower panel) for the parameters in question. The median values of the model parameters reported in the literature for stars are shown by a solid green line. The colored regions represent the measurement errors obtained in this study for the model parameters under consideration.}
    \label{fig:A1}
\end{figure*}

\end{document}